\documentclass{aa}
\usepackage{epsf}
 
\begin{document} 
 
  \thesaurus{12.          
              (08.14.1;   
               02.04.1)   
             } 
 
\newcommand{\pF}{\mbox{$p_{\mbox{\raisebox{-0.3ex}{\scriptsize F}}}$}}  
\newcommand{\vph}[1]{\mbox{$\vphantom{#1}$}}  
\newcommand{\kB}{\mbox{$k_{\rm B}$}}           
\newcommand{\vF}{\mbox{$v_{\mbox{\raisebox{-0.3ex}{\scriptsize F}}}$}}  
\renewcommand{\arraystretch}{1.5} 
 
\title{ Bulk viscosity in superfluid neutron star cores. I.
        Direct Urca processes in  $npe\mu$ matter}
\author{ 
        P.~Haensel\inst{1}\thanks{%
E-mail: haensel@camk.edu.pl}\and K.P.~Levenfish\inst{2}\and D.G.~Yakovlev\inst{2} 
        } 
\institute{ 
 N.~Copernicus Astronomical Center, 
       Bartycka 18, 00-716 Warszawa, Poland 
\and 
        Ioffe Physical Technical Institute, Politekhnicheskaya 26, 
        194021 St.-Petersburg, Russia 
              } 
 
\date{Accepted 28th March 2000} 
 
\offprints{P.~Haensel} 
 
\titlerunning{ Bulk viscosity in superfluid neutron star cores} 
\authorrunning{P.~Haensel et al.} 
 
\maketitle

 
\begin{abstract} 
The bulk viscosity of the neutron star matter due to 
the direct Urca processes
involving nucleons, electrons and muons
is studied taking into account 
possible superfluidity of nucleons in the 
neutron star cores.
The cases of singlet-state pairing or triplet-state 
pairing (without and with nodes of the superfluid gap 
at the Fermi surface) of nucleons are considered.
It is shown that the superfluidity may strongly 
reduce the bulk viscosity. 
The practical expressions for the superfluid reduction 
factors are obtained. 
For illustration, the bulk viscosity is calculated for two 
models of dense
matter composed of neutrons, protons, electrons and muons. 
The presence of muons affects the bulk viscosity 
due to the direct Urca reactions involving electrons 
and produces additional comparable contribution due to 
the direct Urca reactions involving muons. 
The results can be useful for 
studying damping of vibrations of neutron stars 
with superfluid cores. 
\end{abstract} 
 
\section{Introduction} 
 
The dissipative processes in neutron stars play an important 
role for some dynamical properties of these unique objects. 
Shear viscosity  damps differential rotation of neutron stars, 
leading to their uniform rigid-body rotation.  
Quite generally, the viscosity of neutron star matter implies 
damping
of pulsations of neutron stars.
Such 
pulsations could be excited during the process of neutron star 
formation. They could also be triggered by 
instabilities appearing during neutron star evolution,
or by some external perturbations.
The corresponding damping timescales involve  the shear 
and bulk viscosities of neutron star interior. Both 
viscosities depend on density, temperature and composition of 
dense matter. Calculations of damping timescales of
pulsations for various models of neutron star 
interiors have been done
by Cutler et al.\ (\cite{cls90}).
Viscous  damping of pulsations of newly born
hot neutron stars
turns out to be due to the bulk viscosity.

Another role of the viscosity of neutron star matter
is that it can damp
gravitational radiation driven instabilities
in rotating neutron stars and,
therefore, could be important for determination of the
maximum rotation frequency of neutron stars.
In the absence of  viscosity all  rotating neutron stars
would be driven unstable by the emission of gravitational
waves. Viscous damping timescales enter explicitly
the criteria for the appearance  of these  instabilities. 
Similarly, as in pulsating non-rotating neutron stars,
viscous damping of gravitational radiation driven instabilities
in rapidly rotating newly born neutron stars
is dominated by bulk viscosity of neutron star interiors
(e.g., Lindblom \cite{l95}, Zdunik 1996, Lindblom et al.\ \cite{lom98}).

In this paper, we focus on the viscosity
of matter in the neutron star cores 
(which extend from the layers of density 
$\rho \simeq 1.5 \times 10^{14}$ g cm$^{-3}$ 
to the stellar centers). 
It is well known that 
the cores consist of baryons (neutrons $n$, protons $p$ 
and possibly hyperons)
and leptons (electrons $e$ and muons $\mu$). 
All constituents of matter are strongly degenerate fermions. 
The electrons and muons form 
almost ideal Fermi gases. The electrons are ultrarelativistic 
while the muons 
may be non-relativistic or relativistic
depending on density.
The nucleons are, to a good
approximation, non-relativistic and constitute 
strongly interacting Fermi liquid.
At the densities close to the normal nuclear density (baryon 
number density $n_0=0.16$ fm$^{-3}$ which corresponds to the mass density 
$\rho_0=2.8 \times 10^{14}$ g cm$^{-3}$), neutron star matter 
is composed of $n$, $p$, $e$, and $\mu$. At higher 
densities [$\rho \!\ga(3$--$4)\,\rho_0$]
some models of dense matter predict 
appearance of hyperons. At still higher densities, 
some calculations indicate 
possible presence of exotic phases 
(pion condensate, kaon condensate, deconfined quark 
matter). We will not consider the hyperonic or exotic phases
but restrict ourselves to the study of the  
$npe\mu$ matter.
 
Our analysis is 
additionally complicated by possible 
superfluidity of nucleons in the neutron star cores.
The superfluidity is thought to be produced by 
Cooper pairing of nucleons due to attractive parts
of nucleon-nucleon interaction. The superfluidity
of nucleons in the neutron star cores
has been the subject of numerous papers 
(as reviewed, for instance, by  Yakovlev et al.\  \cite{yls99}). 
Various microscopic theories predict very different 
superfluid gaps (critical temperatures $T_{cn}$ and $T_{cp}$) 
of neutrons and protons depending on specific model 
of strong interaction employed 
and specific many-body
theory used to account for medium effects. 
However, all these results have important common features.
In particular, the proton pairing occurs mainly 
in the $^1$S$_0$-state since the $pp$ interaction 
is attractive in this state everywhere in the 
neutron star core due to not too high number density 
of protons. As for the neutrons, their interaction 
in the $^1$S$_0$ state turns from attraction to repulsion at densities 
$\rho \ga \rho_0$ but the interaction in the 
$^3$P$_2$ state may be attractive and may lead to Cooper
pairing. The critical temperatures $T_{cn}$ and $T_{cp}$ 
in the neutron star cores 
predicted by different microscopic theories depend on density 
and scatter in a wide range from about
$10^8$ to $10^{10}$ K. 
Under these conditions we will not rely on any
specific microscopic theory of nucleon superfluidity, 
but will treat $T_{cn}$ and $T_{cp}$ as free parameters.

The viscosity, we are interested in, 
is well known to consist of the shear viscosity 
and bulk viscosity. 
The standard source of the shear viscosity of the neutron star matter 
is scattering between its constituents. Classical 
calculations of shear viscosity for the $npe$ model of 
non-superfluid matter were done by 
Flowers \& Itoh (\cite{fi79}). Their 
results were used in the studies of damping of neutron star 
pulsations by Cutler et al.\ (\cite{cls90}).
In the superfluid core of a rotating neutron star,
there is an additional viscous mechanism, 
called mutual friction,  resulting from
the scattering of electrons off the magnetic 
field trapped in the cores of superfluid
neutron vortices (Lindblom \& Mendell \cite{lm95}).

The bulk viscosity may partly be determined by particle
scattering. However, this component
of bulk viscosity is usually much smaller than the shear viscosity
(e.g., Baym \& Pethick \cite{bp91}).
The main contribution
into the bulk viscosity of sufficiently
hot $npe\mu$ matter comes from the neutrino 
processes of Urca type associated with 
electron and muon emission
and capture by nucleons.
We will focus on such bulk viscosity.
Generally, the neutrino
processes in question
are divided into the {\it direct Urca} and the {\it modified Urca}
processes. A direct Urca process is a sequence of two reactions
(direct and inverse one) and can be written as 
\begin{equation} 
    n \to p + l + \bar{\nu}_l, \quad p + l \to n + \nu_l,
\label{baryon-Durca} 
\end{equation} 
where $l$ is either electron or muon, and $\nu_l$ is an associated
neutrino. The
most important is  the process
(Lattimer et al.\ \cite{lpph91}) involving electrons
($l=e$); it consist of the beta decay
of neutron and subsequent beta capture. It should 
be emphasized that the both direct Urca processes are
forbidden by momentum conservation of reacting particles 
for the simplest model of dense matter 
as a gas of noninteracting Fermi particles (e.g., Shapiro 
\& Teukolsky \cite{st83}) at any density $\rho$ in the neutron star cores. 
Nevertheless they are allowed 
(Lattimer et al.\ \cite{lpph91}) for 
some realistic equations of state at densities
higher than some threshold densities (of 
several $\rho_0$). 
Thus, the direct Urca processes may be open 
in the inner cores of rather massive neutron stars. 
The threshold density for the muon process is always
higher than for the electron one.

If allowed, the direct Urca processes produce the most 
powerful neutrino emission from the neutron star cores 
(Lattimer et al. \cite{lpph91}). Corresponding neutrino emissivities 
were calculated by Lattimer et al.\ (\cite{lpph91})
and used in numerous simulations
of the neutron star cooling as reviewed, for instance, 
by Yakovlev et al.\ (\cite{yls99}). In the absence of nucleon
superfluidity, the direct Urca processes lead to the 
{\it fast} cooling of neutron stars. If allowed, the direct Urca
processes produce the main contribution into the bulk 
viscosity we are interested in. 

However, for many equations of state the direct
Urca processes are forbidden by momentum conservation
at any density in the neutron star cores. Moreover,
they are prohibited at $\rho\la 3\, \rho_0$
for the majority of equations of state.
In such cases, they do not operate in the low and
medium-mass neutron stars and in the outer
cores of all neutron star 
models constructed using these equations of state.
If so, the bulk viscosity
is determined by the
the reactions of the modified 
Urca processes 
\begin{equation} 
    n + N \to p + N + l + \bar{\nu}_l, \quad p + N + l \to n + N + \nu_l,
\label{baryon-Murca} 
\end{equation} 
where $N$ is an additional nucleon required to
conserve momentum of the reacting particles. 
For instance, in $npe$ matter one has two 
modified Urca processes corresponding to 
$N=n$ and
$N=p$, respectively,
which can be referred to as the neutron and proton 
branches of the modified Urca process (e.g., Friman 
\& Maxwell \cite{fm79}, Yakovlev \& Levenfish \cite{yl95}). 
The rates of the modified Urca processes are 
typically 3--5 orders of magnitude lower 
than the rates of the direct Urca processes. 
The modified Urca processes either have no 
density threshold (as the neutron branch in $npe$ matter) 
or have much lower density thresholds 
than the direct Urca processes. Thus they 
operate in the entire neutron star core. If 
the direct Urca processes are forbidden 
and matter is non-superfluid, the modified Urca 
processes produce the main neutrino emission 
from the neutron star cores leading to {\it slow} 
({\it standard}) cooling of neutron stars. 
Their role in the neutron star cooling theory 
has been studied in many papers (see, e.g., 
Yakovlev et al.\ \cite{yls99}, for review). 
 
Thus, the problem of calculating the 
bulk viscosity due to neutrino processes 
is quite complicated: there are several
neutrino processes involved 
influenced by
possible nucleon superfluidity.
So far the bulk viscosity has been studied 
only for non-superfluid $npe$ matter. The viscosity 
due to the neutron branch of the modified Urca process was 
analyzed by Sawyer (\cite{s89}) while the viscosity 
produced by the nucleon direct Urca process
was considered by Haensel \& Schaeffer (\cite{hs92}). 
The effects of superfluidity 
have not been analyzed for the problem of bulk viscosity 
but studied thoroughly for the 
neutrino emissivity produced in different reactions 
(e.g., Yakovlev et al.\ \cite{yls99}, and references therein). 
 
The relative importance of the bulk viscosity produced 
by neutrino reactions with respect to the shear viscosity
produced by collisions can be estimated by comparing 
the results by Sawyer (\cite{s89}) and Haensel \& Schaeffer 
(\cite{hs92}) with the values of the shear viscosity 
calculated by Flowers \& Itoh (\cite{fi79}). The comparison 
shows that the neutrino bulk viscosity dominates 
in the neutron star cores for temperatures $T \ga 10^8$ K 
if the direct Urca processes are switched on and 
for $T \ga 10^9$ K if the direct Urca processes are forbidden. 
In superfluid matter, the bulk viscosity can be even more 
important. 
 
In this paper, we consider the bulk viscosity produced by 
the direct Urca processes
in $npe\mu$ matter of the neutron star
cores. In analogy with the effects of superfluidity 
on the neutrino emissivity, we will analyze
the effects of superfluidity of nucleons on the bulk viscosity.
 
\section{Bulk viscosity in non-superfluid matter} 
\label{sect-bulk} 
 
\subsection{Bulk viscosity in $npe\mu$ matter} 
 
Consider
the bulk viscosity
produced by the direct Urca process (involving muons and
electrons) in non-superfluid $npe\mu$ matter. 

Due to very frequent collisions between particles, 
dense stellar matter 
should very quickly (instantaneously on macroscopic time scales) 
achieve a quasi-equilibrium state 
with certain temperature $T$ 
and chemical potentials $\mu_i$ of different particle species 
$i=n,p,e,\mu$. 
Typically, 
all particle species 
are strongly degenerate. 
We assume that the matter is transparent for neutrinos, which 
therefore do not contribute to the thermodynamical quantities. 
 
A quasi-equilibrium state described above does not 
mean full thermodynamic equilibrium. The latter 
assumes additionally the equilibrium with respect to the beta 
and muon decay 
and capture processes. We will call it the {\it chemical equilibrium}. 
Relaxation to the chemical equilibrium depends drastically 
on a given equation of state and local density of matter $\rho$. 
It is realized either through direct Urca or modified 
Urca processes (Sect.\ 1). Consideration of this subsection 
is valid for all Urca processes although 
the practical expressions (Sects.\ 2.3 -- 2.4) will
be obtained for the direct Urca processes. 
 
The Urca processes of both types, 
direct and modified, are rather slow. 
Although 
the chemical relaxation rate depends strongly on temperature, in any case 
it takes much more time 
(from tens of seconds to much longer time intervals) 
than the rapid relaxation to a quasi-equilibrium state described above. 
Therefore, a neutron star can be in a quasi-equilibrium, 
but not in the chemical equilibrium, for a long time. 
 
If the chemical equilibrium is achieved, then the chemical 
potentials satisfy the equalities 
$\mu_n=\mu_p+\mu_e$ and 
$\mu_n=\mu_p+\mu_\mu$, which imply 
$\mu_e=\mu_\mu$. 
Under these conditions, the rates 
of the direct and inverse 
reactions, $\Gamma_l$ and $\bar{\Gamma}_l$ ($l$ = $e$ or $\mu$), 
of any Urca process are equal. 
 
Let us assume that the neutron star 
undergoes
radial pulsations of frequency $\omega$. 
Associated temporal variations of the local baryon 
number density will be taken in the form 
$n_b=n_{b0} + n_{b1} 
\cos \omega t$, 
where $n_{b1}$ is the pulsation amplitude 
and $n_{b0}$ is the 
non-perturbed 
baryon number density ($|n_{b1}| \ll n_{b0}$). 
We assume further that $n_{b0}$ corresponds to the chemical equilibrium. 
This chemical equilibrium is violated slightly 
in pulsating matter. 
If the pulsation frequency $\omega$ 
were much smaller than 
the chemical relaxation rates,
the composition of matter 
would follow instantaneous values of $n_b$, 
realizing the chemical equilibrium every moment of time. 

In reality, 
the typical frequencies of the fundamental mode of 
the radial pulsations 
$\omega \sim 10^3$--$10^4$~s$^{-1}$, are 
much higher than the chemical relaxation rates.
As a result, 
the partial fractions $X_i=n_i/n_b$ of all the constituents 
of dense matter are almost unaffected by 
pulsations (i.e., almost constant).
Owing to the slowness of
the Urca reactions, these fractions lag
behind their instantaneous equilibrium values, 
producing non-zero differences of instantaneous $\mu_i$:
\begin{equation} 
       \eta_e=\mu_n-\mu_p-\mu_e, \quad 
       \eta_\mu=\mu_n-\mu_p-\mu_\mu. 
\label{mu} 
\end{equation} 
This causes an asymmetry of the direct and 
inverse direct Urca reactions, and, 
hence, slight deviations from the chemical equilibrium. The asymmetry, 
calculated in the {\it linear approximation} 
with respect to $\eta_l$, is given by 
\begin{equation} 
       \Gamma_l  - \bar{\Gamma}_l 
       = - \lambda_l \,\eta_l \, , 
\label{lambda} 
\end{equation} 
where $\lambda_l$ are the coefficients specified 
in Sect.\ 2.4 for the direct Urca reactions. 
Microscopic calculation (Sect.\ 2.4) yields 
$\lambda_e = \lambda_\mu$. 
In this paper, we will restrict ourselves to the case
$|\eta_l| \ll T$ ($l= e,\, \mu$).
Our definition of $\lambda_l$ is the same as 
was used by Sawyer (\cite{s89}) for the case of $npe$ matter. 
Notice that $\lambda_l$ defined in this way is negative. 
 
The non-equilibrium Urca reactions 
provide the energy dissipation 
which causes damping of stellar pulsations. 
Accordingly, they contribute to the bulk viscosity
of matter, $\zeta$. Using the standard 
definition of the bulk viscosity, 
the energy dissipation rate per unit volume averaged over 
the pulsation period ${\cal P}=2\pi/\omega$ can be written as 
\begin{equation} 
      \left\langle  \dot{\cal E}_{\rm kin} \right\rangle 
      = - \, {\zeta \over 
      \cal P} \,\int_0^{\cal P} \!\! {\rm d}  t 
      \left( {\rm div}\, {\vec v} \right)^2 
      = - { \zeta \, \omega^2 \over  2} \, 
      \left( \frac{n_{b1}}{n_{b0}} \right)^2 \, , 
\label{dEkin} 
\end{equation} 
where ${\vec v}$ is the hydrodynamic velocity associated 
with the pulsations. The latter equality is obtained from 
continuity equation for baryons, $ \dot{n}_b + 
n_{b0} \, {\rm div} {\vec v} = 0$ (pulsations do not 
change their total number), which yields 
$\: {\rm div} \, {\vec v} = 
-\dot{n_b}/n_{b0}= \omega \,(n_{b1}/n_{b0}) \, \sin \omega t$.

The hydrodynamic matter flow implied by the stellar pulsations is 
accompanied by the time variations of the local pressure, 
$P(t)$. The dissipation of the energy of the hydrodynamic motion 
is due to irreversibility of the periodic compression-decompression 
process. Averaged over the pulsation period, this dissipation rate in 
the unit volume is 
\begin{equation} 
     \left\langle \dot{\cal E}_{\rm diss} \right\rangle = 
     - \frac{n_b}{\cal P} \int_0^{\cal P} \!\! {\rm d} t \; 
        P \,  \dot{V}. 
\label{dEdiss} 
\end{equation} 
For a strictly reversible process, 
     $\left\langle \dot{\cal E}_{\rm diss} \right\rangle =0$. 
However, in our case 
the quantities $P$ and $V$ 
follow variations of $n_b$ in different ways. 
The specific volume $V=1/n_b$ varies instantaneously as $n_b$ varies, 
i.e., the oscillations of $V$ and $n_b$ are in phase 
but the pressure varies with certain phase shift. In $npe\mu$ matter 
at quasi-equilibrium the pressure can be regarded 
as a function of four variables: $n_b$, $X_e$, $X_\mu$, and $T$. 
Variations of $T$ are insignificant, for our problem, 
and may be disregarded. 
Thus, it is sufficient to assume that $P=P(n_b, X_e, X_\mu)$. 
Variations of the pressure contain the 
terms oscillating with shifted phases 
due to the lags of $X_e$ and $X_\mu$. 
 
Let us evaluate the integral (\ref{dEdiss}). 
We have 
$\dot{V}= - \dot{n}_b/n_{b0}^2 
= \omega \, (n_{b1}/n_{b0}^2) \sin \omega t $. Thus the 
only terms in $P$ contributing 
into the energy dissipation are 
those which are proportional to $\sin \omega t$. 
At this stage it is convenient 
to use the formalism of complex variables and 
write $P=P_0 + {\rm Re} \{ P_1 \, \exp( i \omega t) \}$, 
$X_l=X_{l0}  + {\rm Re} \{ X_{l1} \, \exp (i \omega t) \}$, 
where $P_0$ and $X_{l0}$ are the equilibrium quantities while 
$P_1$ and $X_{l1}$ are 
small complex amplitudes to be determined. 
We have 
\begin{equation} 
   P_1 = 
         \left( \frac{\partial P}{\partial n_b} \right) n_{b1} 
         + \left( \frac{\partial P}{\partial X_e} \right) X_{e1} 
         + \left( \frac{\partial P}{\partial X_\mu} \right) X_{\mu1} , 
\label{P} 
\end{equation} 
where all the derivatives are taken at equilibrium. 
The real part of $P$ contains the terms with 
$\sin \omega t$ provided the amplitudes $X_{l1}$ 
have imaginary part. 
 
The change of the lepton fraction 
$\dot{X}_l$ is determined by the difference of the direct and inverse 
reaction rates  
given by Eq.\ (\ref{lambda}). 
The quantity $\eta_l$ in the latter equation varies 
near its equilibrium value $\eta_{l0}=0$ as 
$\eta_l = \eta_{l0} + 
{\rm Re} \{ \eta_{l1} \exp(i\omega t)\}$, 
where 
\begin{equation} 
 \eta_{l1} = 
   \left( \frac{\partial \eta_l}{\partial n_b} \right) n_{b1} 
  + \left( \frac{\partial \eta_l}{\partial X_e} \right) X_{e1} 
  + \left( \frac{\partial \eta_l}{\partial X_\mu} \right) X_{\mu1}, 
\label{dmu} 
\end{equation} 
and all the derivatives are again taken at equilibrium. 
Combining the expression 
$n_{b0} \dot{X}_l=\Gamma_l - \bar{\Gamma}_l$ 
with Eq.\ (\ref{lambda}) and using the formalism 
of complex variables we obtain the two equations
$X_{l1}= -\lambda_l \, \eta_{l1}/(i\omega n_b)$ 
(for $l=e$ and $\mu$). These two equations supplemented by 
Eq.\ (\ref{dmu}) constitute a system of equations 
which solution is 
\begin{equation} 
  X_{l1} = - { n_{b1} \over n_{b0}} \, 
             { C_l \,(B_{l^\prime l^\prime}+ i\, \alpha_{l^\prime}) 
             - C_{l^\prime} \, B_{l l^\prime} 
             \over 
             (B_{ll}+ i\, \alpha_l) 
             (B_{l^\prime l^\prime}+ i\, \alpha_{l^\prime}) 
              - B_{l^\prime l} \, B_{l l^\prime}} , 
\label{complex} 
\end{equation} 
where $l' \neq l$ and $\alpha_l=\omega \,n_{b0}/\lambda_l$. 
In analogy with Sawyer (\cite{s89}) we
have introduced the notations: 
\begin{equation} 
     B_{ll^\prime}= 
         \frac{\partial \eta_l }{\partial  X_{l^{\prime}} }, \quad 
      C_l = n_{b0} \,\frac{\partial \eta_l }{\partial n_b }. 
\label{coef} 
\end{equation} 
Note that all the derivatives are taken at equilibrium. 
In the absence of muons from 
Eq.\ (\ref{complex}) we have $X_{\mu 1} \equiv 0$ 
and $X_{e1} = - (n_{b1}/n_{b0})\,C_e/(B_{ee}+i \alpha_e)$. 
This is the well known limit considered by 
Sawyer (\cite{s89}) and Haensel \& Schaeffer (\cite{hs92}). 
 
Generally, 
Eq.\ (\ref{complex}) is quite complicated. 
However, in practical applications stellar oscillations are always much 
more frequent than the beta and muon reaction rates 
($|\partial \eta_l/\partial X_l| \ll \omega n_{b0}/ |\lambda_l| $) 
and it is sufficient to use the asymptotic 
form of the solution in the {\it high-frequency limit}. In this limit 
the imaginary part of $X_{l1}$ is related to the amplitude $n_{b1}$ as 
\begin{equation} 
       {\rm Im} \left\{ X_{l1}  \right\} 
        =   \frac{n_{b1}}{n_{b0}}\, \frac{\lambda_l}{\omega }\, 
            \frac{\partial \eta_l }{\partial n_b } \, . 
\label{dX1} 
\end{equation} 
Combining this equation with that for $\dot{V}$ 
(see above) and inserting into Eq.\ (\ref{dEdiss}) we get 
the dissipation rate of mechanical energy 
\begin{equation} 
     \left\langle \dot{\cal E}_{\rm diss}  \right\rangle = 
     \frac{\omega^2}{2} \, 
     \left( \frac{n_{b1}}{n_{b0}} \right)^2 \, 
     \sum_l \, 
     \frac{\lambda_l}{\omega^2} \: \frac{\partial P}{\partial X_l} \: 
     \frac{\partial \eta_l }{\partial n_b }. 
\label{dEdiss1} 
\end{equation} 

Finally, bearing in mind that 
$ \langle \dot{\cal E}_{\rm kin} \rangle = - 
  \langle \dot{\cal E}_{\rm diss} \rangle $, 
from Eqs.\ (\ref{dEkin}) and (\ref{dEdiss1}) 
we have the bulk viscosity 
\begin{equation} 
    \zeta = \zeta_e + \zeta_\mu \, , \quad 
    \zeta_l = \frac{ | \lambda_l | }{\omega^2} \: 
              \left| \frac{\partial P}{\partial X_l} \right| \: 
              \frac{\partial \eta_l }{\partial n_b }. 
\label{zeta} 
\end{equation} 
Here we have taken into account that $\lambda_l$ and 
$\partial P /\partial X_l$ are negative and presented the viscosity 
in the form which clearly shows that $\zeta_l$ is positive. 
The expression for $\zeta_e$ was obtained by Sawyer (\cite{s89}).
Let us emphasize that the viscosity we deal with
has meaning of a  coefficient in the equation
that determines the damping rate of stellar pulsations
averaged over pulsation period (and it cannot generally be used
in exact hydrodynamical equations of fluid motion). 

Therefore, in the high frequency limit, 
which is the most important in practice,
the bulk viscosity $\zeta$ is a sum of the 
partial viscosities $\zeta_e$ and $\zeta_\mu$ 
produced by the electron and muon Urca processes, respectively. 
This additivity rule greatly simplifies evaluation of $\zeta$. 
The values of $\partial P/\partial X_l$ and 
$\partial \eta_l / \partial n_b $ are determined 
by an equation of state as described 
in Sects.\ 2.2 and 2.3. The factors 
$\lambda_l$ are studied in Sect.\ 2.4 for the direct Urca reactions.
The results of analogous consideration for the modified 
Urca reactions will be published elsewhere.

\subsection{Partial bulk viscosity} 
\label{Bulk} 
 
Let us discuss briefly how to calculate 
the partial bulk viscosity $\zeta_l$ of $npe\mu$ matter 
for a given equation of state. 
All the quantities in this section and below are 
essentially (quasi)equilibrium values. Thus we will omit 
the index $0$, for brevity.
 
Since the electrons and muons constitute almost 
ideal gases, the 
matter  energy
per baryon can be generally written as 
\begin{equation} 
   E=E_N(n_b,X_p)+X_e E_e(n_e)+ X_\mu E_\mu(n_\mu), 
\label{Helmholtz} 
\end{equation} 
where $E_N(n_b,X_p)$ is the 
nucleon energy
per baryon, 
$X_p=n_p/n_b$ is the proton fraction, 
and $E_l(n_l)$ is a lepton energy per one lepton ($e$ or $\mu$). 
The latter energy is determined by the lepton number density, $n_l$. 
Owing to charge neutrality, we have $X_p=X_e+X_\mu$. 
 
The neutron and proton chemical potentials are 
given by $\mu_n=\partial (n_b E_N) / \partial n_n$ 
and $\mu_p=\partial (n_b E_N) / \partial n_p$. 
The derivatives should be taken using  
$n_p=X_p \, n_b$ and $n_n=(1-X_p) \, n_b$. 
This gives $\mu_n - \mu_p = - \partial E_N / \partial X_p$. 
The chemical potentials of electrons or muons are 
$\mu_l= (m_l^2 c^4 + p_{{\rm F}l}^2 c^2)^{1/2}$, 
where $p_{{\rm F}l}=\hbar \left( 3\pi^2 \,n_l \right)^{1/3}$ 
is the Fermi momentum. Therefore, the difference of chemical potentials is 
\begin{equation} 
   \eta_l= - {\partial \,E_N(n_b,X_p) \over \partial X_p} - \mu_l. 
\label{eta_l} 
\end{equation} 

Now let us calculate $C_l$ from Eq.\ (\ref{coef}).
The derivative of $\mu_l$ with respect to $n_b$ 
is evaluated using  $n_l=X_l  n_b$. The result is 
\begin{equation} 
   C_l = - n_b \; { \partial^2 E_N(n_b,X_p) 
          \over \partial n_b \, \partial X_p} 
           - {c^2 p_{{\rm F}l}^2 \over 3\, \mu_l}. 
\label{Cl1} 
\end{equation} 
Using Eq.\ (\ref{Helmholtz}) and the standard thermodynamic 
relations we obtain the pressure $P=P_N+P_e+P_\mu$, 
where $P_N=n_b^2 \, \partial E_N / \partial n_b$ is 
the nucleon pressure, while $P_e$ and $P_\mu$ are 
the well known partial pressures of free gases of $e$ and $\mu$, 
respectively. Direct calculations yields 
$\partial P /\partial X_l= - n_b \, C_l$. Inserting 
this derivative into Eq.\ (\ref{zeta}) 
and using the definition of $C_l$ 
we come to a very simple equation 
\begin{equation} 
       \zeta_l=  \frac{|\lambda_l|}{\omega^2}\: C_l^2 . 
\label{zeta_C} 
\end{equation} 
Thus, a partial bulk viscosity $\zeta_l$ 
is expressed through the two factors, $C_l$ and $\lambda_l$. 
Calculation of $C_l$ is discussed in Sect.\ 2.3, while 
$\lambda_l$ is analyzed in Sect.\ 2.4. 
 
\subsection{Illustrative model of $npe\mu$ matter} 
\label{EOS} 
 
For illustration, we use a phenomenological equation of state 
proposed by Prakash et al.\ (\cite{pal88}). 
According to these authors the nucleon energy is 
presented in the familiar form 
(neglecting small neutron-proton mass difference)
\begin{equation} 
   E_N(n_b,X_p) = E_{N0}(n_b) + S(n_b)\, (1- 2X_p)^2 , 
\label{E_N} 
\end{equation} 
where 
$E_{N0}(n_b)=E_N (n_b,X_p=1/2)$ is the energy of the symmetric 
nuclear matter and  $S(n_b)$ is the symmetry energy. 
From Eq.\ (\ref{eta_l}) at equilibrium ($\eta=0$) 
we immediately obtain $\mu_l=4(1-2X_p)S(n_b)$, and 
from Eq.\ (\ref{Cl1}) we have 
\begin{eqnarray} 
   C_l & =  & 4 \, (1-2X_p) \, n_b \, {{\rm d} S \over {\rm d}n_b} 
           - {c^2 p_{{\rm F}l}^2 \over 3\, \mu_l}
\nonumber \\
     & =  &  (1-2X_p) \, n_b \, {{\rm d}~~~  \over {\rm d}n_b} \,
            \left( { \mu_l \over  1-2X_p} \right) \,   
           - {c^2 p_{{\rm F}l}^2 \over 3\, \mu_l}. 
\label{C_emu} 
\end{eqnarray} 
This is the practical expression for evaluating $C_l$. 
The factor $C_l$ is not affected by a 
possible nucleon superfluidity which has a negligible 
effect on the equation of state. 
The relative effect of the superfluidity 
on the energy per nucleon is $\sim (\Delta /\mu_N)^2 \sim 
10^{-4}$--$10^{-3}$, where $\Delta$ is the superfluid energy gap, 
and $\mu_N$ is the nucleon Fermi energy. 
 
In accordance with Eqs.\ (\ref{C_emu}) and (\ref{zeta_C}), 
the bulk viscosity is determined by 
the symmetry energy $S(n_b)$. 
At the saturation density $n_0=0.16$~fm$^{-3}$ 
the symmetry energy 
$S_0=S(n_0)$ is measured rather reliably in laboratory 
(e.g., Moeller et al.\ \cite{mmst88}) 
but at higher $n_b$ it is still unknown. 
Prakash et al.\ (\cite{pal88}) 
presented $S(n_b)$ in the form: 
\begin{equation} 
   S(n_b) = 13\, {\rm MeV} \left[ u^{2/3}-F(u) \right] +S_0  F(u), 
\label{S} 
\end{equation} 
where $u=n_b/n_0$, $S_0=30$~MeV, and $F(u)$ satisfies the condition 
$F(1)=1$. They proposed three 
theoretical models (I, II and III) for $F(u)$: 
\begin{equation} 
   F_{\rm I}(u)=u, \quad F_{\rm II}(u)= {2 \, u^2 \over u + 1}, \quad 
   F_{\rm III}(u)=\sqrt{u}, 
\label{FFF} 
\end{equation} 
and three models for $E_{N0}(n_b)$ in Eq.\ (\ref{E_N}). 
We do not discuss the latter models here because they are not 
required to calculate the particle fractions and 
the bulk viscosity as a function of $n_b$. 
Following Sawyer (\cite{s89}) and Haensel \& Schaeffer (\cite{hs92}) 
we will use models I and II, for illustration. 
Model I gives lower symmetry energy and accordingly lower 
excess of neutrons over protons. 
In contrast to the above authors we will
allow for appearance of muons. 
 
The three models for $E_{N0}(n_b)$ 
correspond to three different values of the compression modulus 
of symmetric nuclear matter at saturation, $K_0$=120, 180 and 
240 MeV. If, for instance, we take models I and II 
of $S(n_b)$ and the  model of $E_{N0}(n_b)$ with 
$K_0=180$ MeV, we have two  model equations 
of state of matter (models I and II) 
in the cores of neutron stars. 
The effective masses of nucleons, 
renormalized by the medium effects, will be set equal 
to 0.7 of their bare masses (the same values will be 
adopted in all numerical examples below). 
The equations of state I and II obtained in this way are moderately stiff. 
The maximum neutron star masses for 
models I and II are 
$M_{\rm max} = 1.72 \, M_\odot$ and $1.74 \, M_\odot$, respectively. 
 
The equilibrium fractions of muons and electrons, 
$X_\mu$ and $X_e$, can be obtained 
as numerical solutions of 
the set of the chemical equilibrium equations at given $n_b$: 
\begin{eqnarray} 
&&   X_\mu  = {1 \over 2} - {\cal A} \, X_e^{1/3} - X_e , 
\nonumber \\ 
&&   X_e^{2/3} - X_\mu^{2/3} - 
         {m_\mu^2 \, c^2 \over 
          \hbar^2  \left( 3\pi^2 n_b  \right)^{2/3} } = 0 , 
\label{Xemu} 
\end{eqnarray} 
where ${\cal A} = \hbar\, c \, ( 3\pi^2 n_b )^{1/3}/(  8 \, S)$. 
If the muons are absent ($X_\mu = 0 $), 
the second equation should be disregarded, 
while the the first one 
determines the equilibrium composition of matter 
\begin{equation} 
         X_e^{1/3} = \left(  \sqrt{ D }  + \frac{1}{4}  \right)^{1/3} 
          - \left( \sqrt{ D } 
              - \frac{1}{4} 
            \right)^{1/3}  , 
\label{X_0} 
\end{equation} 
where $D=({\cal A}^3/27)+(1/16)$.
At given $n_b$ the equilibrium 
fraction of electrons in $npe\mu$-matter is always smaller than 
it would be in $npe$ matter, while the equilibrium fraction 
of protons is always higher. 
The threshold of muon appearance is 
determined by the condition $\mu_e= m_\mu c^2$. 
For models I and II, the muons appear at the baryon number density 
0.150 fm$^{-3}$ and 0.152 fm$^{-3}$, 
respectively. 
 
In Fig.\  1  
we plot the factors 
$C_e(n_b)$ (solid lines) and 
$C_\mu(n_b)$ (dot-and-dash lines) which determine the 
bulk viscosity for models I and II. The dotted lines show 
$C_e(n_b)$ for the simplified models I and II in which 
appearance of muons is artificially forbidden. 
These results coincide with those obtained by 
Sawyer (\cite{s89}) and Haensel \& Schaeffer (\cite{hs92}). 
They coincide also with the solid lines at densities below 
the thresholds of muon appearance but go above 
the solid lines at higher densities (the 
presence of muons affects fractions of electrons and protons). 
As for the factor $C_\mu(n_b)$, it appears in a 
jump-like manner at the muon threshold, 
 initially
exceeds $C_e(n_b)$ 
 and then
tends to $C_e(n_b)$ with increasing density. 
 
\begin{figure}[t] 
\label{fig:Cl} 
\begin{center} 
\vspace*{-0.6cm} 
\epsfxsize=10.5cm 
\epsfbox{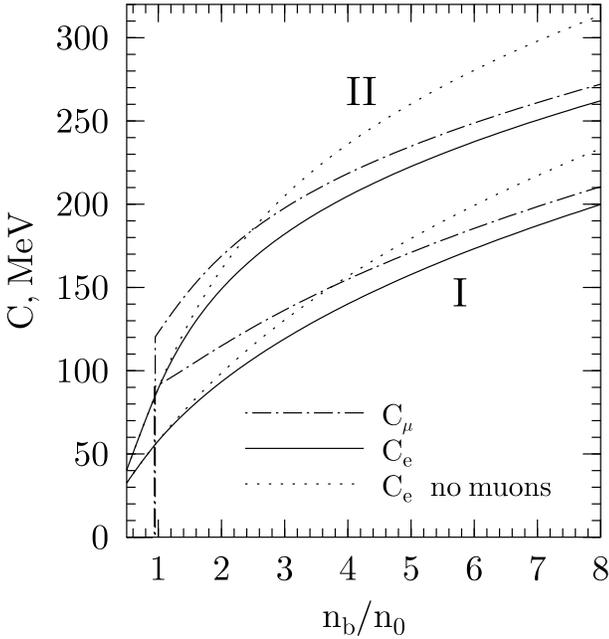} 
\vspace*{-0.5cm} 
\caption[]{
        Linear response of chemical potential 
        difference to baryon number density perturbation, 
        $C_l=n_b\,\partial\eta_l/\partial n_b$, 
        versus the baryon number density $n_b$ 
        for two model equations of state 
        proposed by Prakash et al.\
        (\protect{\cite{pal88}}) and discussed in the text.
        Solid curves correspond to $l=e$, 
        while dot-and-dash curves to $l=\mu$. 
        Dotted lines present $C_e$ for the simplified 
        models of matter without muons. 
                          } 
\end{center} 
\vspace*{-0.7cm} 
\end{figure} 
 
\subsection{Practical expressions for the bulk viscosity of 
            $npe\mu$ matter} 
\label{sect-lambda} 
 
Now let us calculate the factor $\lambda_l$, which 
enters the bulk viscosity (\ref{zeta_C}) and 
determines the asymmetry (\ref{lambda}) of the rates of 
direct and inverse reactions of the direct Urca processes in $npe\mu$ matter. 
In the absence of superfluidity, the rate of a direct reaction 
producing a lepton $l$ is given by ($\hbar=c=\kB =1$) 
\begin{eqnarray} 
     \Gamma_l & = & 
     \int\! \left[ \prod^2_{j=1}  \frac{{\rm d}^3 p_j}{(2\pi)^3} \right] \: 
     \frac{{\rm d}^3 p_l}{2\varepsilon_l (2\pi)^4}\, 
     \frac{{\rm d}^3 p_\nu}{2\varepsilon_\nu (2\pi)^3}\, 
     f_1 (1-f_2)(1-f_e) 
\nonumber \\ 
     & & \times \,
     (2\pi)^3\:\delta(E_f-E_i)\:\delta({\vec P}_f -{\vec P}_i)\, 
     \sum_{\rm spins} |M|^2, 
\label{Gamma_l} 
\end{eqnarray} 
where ${\vec p}_j$ is a nucleon momentum 
($j=1$ or 2), ${\vec p}_l$ and 
$\varepsilon_l$ are, respectively, the lepton momentum and energy, 
${\vec p}_\nu$ and 
$\varepsilon_\nu$ are the neutrino momentum and energy, 
$\delta(E_f-E_i)$ and $\delta({\vec P}_f-{\vec P}_i)$ 
are the delta functions, which conserve 
energy $E$ and momentum ${\vec P}$ of the particles 
in initial ($i$) and final ($f$) states, $|M|^2$ is the squared 
matrix element of the 
reaction, and $f_i$ is an 
appropriate Fermi-Dirac function, 
$f_i=\left\{ 1+\exp\left[ (\varepsilon_i - \mu_i)/T 
 \right] \right\}^{-1}$. 
Equation (\ref{Gamma_l}) 
includes the instantaneous chemical 
potentials $\mu_i$ ($i=n,p,l$) and does not generally require 
the chemical equilibrium. 
 
For further analysis we introduce the dimensionless quantities: 
\begin{equation} 
   x_i = \frac{\varepsilon_i - \mu_i}{ T} \, , \quad 
   x_\nu = \frac{\varepsilon_\nu}{ T} \, , \quad 
   \xi = \frac{ \eta_l }{T} \, , 
\label{DimLess1} 
\end{equation} 
where the chemical potential difference $\eta_l$ is 
determined by Eq.\ (\ref{mu}). 
Thus, the delta function in Eq.\ (\ref{Gamma_l}) takes the form 
$\delta(E_f-E_i)=T^{-1}\, 
\delta(x_n -x_p -x_e -x_\nu +\xi )$, 
where $\xi=0$ for chemical equilibrium. 
 
Multidimensional 
integrals
in
Eq.\ (\ref{Gamma_l}) 
are standard (see, e.g., Shapiro \& Teukolsky \cite{st83}). 
Equation (\ref{Gamma_l}) is simplified taking into account 
that nucleons and leptons $l$ ($e$ and $\mu$) are 
strongly degenerate. 
The main contribution into the integral comes from 
the narrow vicinities of momentum space near 
the Fermi surfaces of these particles. 
The momenta of nucleons and leptons ($e$ or $\mu$) 
can be set equal to their Fermi momenta in all smooth functions. 
The squared matrix element summed over 
the spin states and averaged over orientations of 
the neutrino momenta is 
\begin{equation} 
    \sum_{\rm spins}|M|^2= G^2 \, (f_V^2 + 3 \, g_A^2) . 
\label{matrix-element} 
\end{equation} 
Here, $G=G_{\rm F} \, \cos\theta_{\rm C}$, 
$G_{\rm F} = 1.436 \times 10^{-49} $ erg~cm$^3$ is the Fermi 
weak coupling constant, $f_V \approx 1$ is the vector 
normalization constant, 
$g_A = 1.23$ is the axial vector normalization 
constant, and $\theta_{\rm C}$ is the Cabibbo angle 
($\sin \theta_{\rm C} = 0.231$). Hence the squared 
matrix element is constant and can be 
taken out of the integral. Further procedure consists in the 
standard energy-momentum decomposition of the integration in 
Eq.\ (\ref{Gamma_l}). It yields: 
\begin{eqnarray} 
 \Gamma_l  =  \Gamma_0 \, I\, ; \quad 
        I & = & 
         \int \! {\rm d} x \, x_\nu^2 
         \int \! {\rm d} x_n\,        
         {\rm d} x_p\, {\rm d} x_l\, f(x_n)f(x_p)f(x_l) 
\nonumber \\ 
      & & \times \, \delta\left( x_n + x_p + x_l -x_\nu +\xi \right).
\label{I1} 
\end{eqnarray} 
We have transformed all the blocking factors $(1-f(x))$ into 
the Fermi-Dirac functions $f(x)$ by replacing $x \to -x$. The 
prefactor $\Gamma_0$ is given by 
(returning to the standard physical units) 
\begin{eqnarray} 
  \Gamma_0 & = & 
       \frac{ G^2 \, (1+3g_A^2) 
        }{ 4 \pi^5\, \hbar^{10} \, c^3} \: \: 
          m^\ast_n \, m^\ast_p\, m^\ast_e \, (\kB T)^5 \,\Theta_{npl} 
       = 1.667 \times 10^{32} 
\nonumber \\ 
       & & \times \,
       \frac{m^\ast_n}{m_n} \, 
        \frac{m^\ast_p}{m_p} \, \left( \frac{n_e}{n_0} \right)^{1/3} 
          \! T_9^5 \, \Theta_{npl} \; \;{\rm cm^{-3}\,s^{-1}} . 
\label{Pref} 
\end{eqnarray} 
Here, 
$T_9$ is temperature in units of $10^9$~K; $m_n^\ast$ 
and $m_p^\ast$ are, respectively, the effective masses of neutrons and 
protons in dense matter (which differ from the bare nucleon 
masses due to the in-medium effects).
Moreover, we have defined
$m^\ast_e\equiv \mu_e/c^2 \approx p_{{\rm F}e} /c $
and $m_\mu^\ast \equiv \mu_\mu/c^2 \approx p_{{\rm F}e}/c$,
for leptons.
The step function $\Theta_{npl}$ equals 1 if the direct 
Urca process is switched on and equals 0 otherwise (Sect.\ 1). 
The direct Urca process of study is switched on if the 
Fermi momenta of the reacting particles satisfy the inequality 
$p_{{\rm F}n}<(p_{{\rm F}p}+p_{{\rm F}l})$ 
(Lattimer et al.\ \cite{lpph91}). 
 
It easy to show that the rate $\bar{\Gamma}_l$ 
of the inverse reaction of the direct 
Urca process (lepton capture) 
differs from the rate of the direct reaction, 
given by Eq.\ (\ref{Gamma_l}), 
only by the argument of the delta function in the expression for 
$I$ (one should replace $\xi \to -\xi$ there). 
Therefore, the difference of 
the lepton production and capture rates 
(\ref{lambda}) can be written as 
\begin{eqnarray} 
&&  
  \Gamma_l - \bar{\Gamma}_l = \Gamma_0 \, \Delta I , 
\label{DeltaGamma} \\ 
&&   \Delta I = \int_0^{\infty} \! {\rm d} x_\nu \, x_\nu^2 
       \left[ 
              J(x_\nu-\xi ) - J(x_\nu+\xi) 
       \right], 
\label{dI}  
\end{eqnarray} 
where
\begin{eqnarray} 
 J(x ) & = & 
         \int \! {\rm d} x_n\, 
         {\rm d} x_p\, {\rm d} x_e\, 
         \; f(x_n)f(x_p)f(x_e)\; 
\nonumber \\ 
   & &    \times \, \delta \left( x_n + x_p + x_e - x \right).
\label{Jpm} 
\end{eqnarray} 
In a non-superfluid matter, which we consider in this section, 
the function $J(x)$ 
is calculated analytically 
\begin{equation} 
  J (x) = {\pi^2 + x^2 \over 2 \, ( 1 + {\rm e}^x ) } \, .
\label{Jpm1} 
\end{equation} 

We see that the difference (\ref{DeltaGamma}) 
of the non-equilibrium rates 
of the direct Urca reactions in normal matter 
is determined solely by 
the parameter $\xi = \eta /T$. 
Moreover, the 
integral (\ref{dI}) is taken analytically for any 
$\xi$: 
\begin{equation} 
    \Delta I = \frac{17\pi^4}{60}\, \xi \, {\cal F}(\xi)\, , \;\;\; 
     {\cal F}(\xi) =  1 + \frac{10}{17 \pi^2} \, \xi^2 + 
                             \frac{1}{17 \pi^4} \, \xi^4 \, . 
\label{F_xi} 
\end{equation} 
This relation, with account for Eqs.\ (\ref{lambda}) and (\ref{dI}), 
gives the factor $\lambda$: 
\begin{equation} 
     |\lambda| = \frac{\Gamma_0 }{ T}\; \frac{\Delta I}{\xi}. 
\label{lambda0} 
\end{equation} 
In this paper, 
we do not consider large deviations from the 
chemical equilibrium. 
We  restrict ourselves to 
the deviations 
$|\eta | \ll T$ for which ${\cal F} (\xi) \approx 1$. 

\begin{figure}[t]
\begin{center} 
\vspace*{-1.1cm} 
\epsfxsize=11cm 
\epsfbox{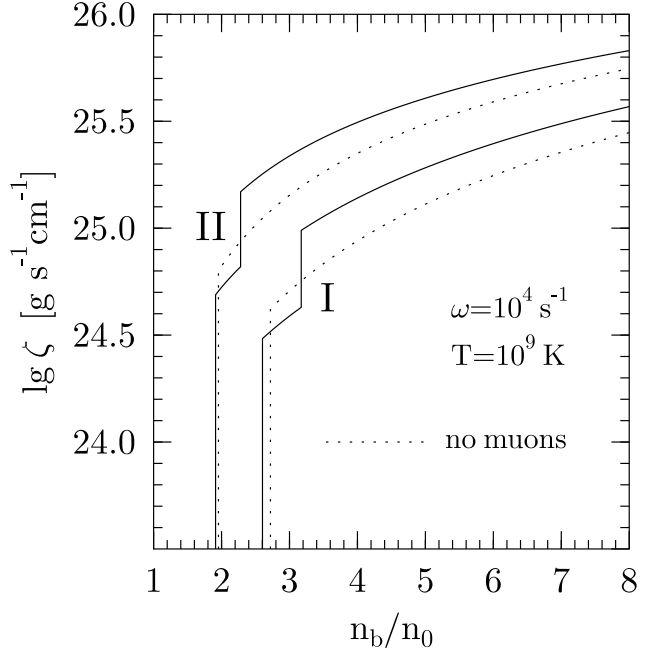} 
\caption[]{\footnotesize 
     The bulk viscosity \protect{$\zeta_0$} for models I and II 
     of non-superfluid $npe\mu$ matter (solid lines) 
     induced by the non-equilibrium direct Urca processes involving 
     electrons and muons 
     versus the baryon number density $n_b$, 
     for $T=10^9$ K and $\omega=10^4$ s$^{-1}$. 
     Dotted lines show the same bulk viscosity but for 
     the simplified models of matter without muons. 
     } 
\end{center}  
\vspace*{-1.1cm}
\label{fig:zeta0} 
\end{figure} 
 
Finally, combining  Eqs.\ (\ref{zeta_C}) and (\ref{lambda0}), 
we obtain the partial bulk viscosity of 
$npe\mu$ matter, 
$\zeta_l=\zeta_{l0}$ (subscript 0 refers to non-superfluid matter), 
induced by a non-equilibrium 
direct Urca process
for $|\eta| \ll T$: 
\begin{eqnarray} 
\zeta_{l0} & = & {17 \, G^2(1+3 g_A^2) C_l^2 \over 
                 240 \pi \hbar^{10} c^3 \, \omega^2 }\, 
                 m^\ast_n m^\ast_p m^\ast_e \,(k_{\rm B}T)^4 \, \ 
                 \Theta_{npl} 
\nonumber \\ 
         &= & 8.553 \times 10^{24} \; \frac{m^\ast_n}{m_n} \, 
      \frac{m^\ast_p}{m_p}\: 
      \left( \frac{n_e}{n_0} \right)^{1/3}  
\nonumber \\ 
        & &  \times \,
      T_9^4 \, \frac{1}{\omega^2_4} \, 
      \left( \frac{C_l}{\mbox{100 MeV} } \right)^2 \, \Theta_{npl} \: 
      \, {\rm g \,cm^{-1}\,s^{-1}} , 
\label{zeta_NSF} 
\end{eqnarray} 
where $\omega_4 = \omega /(10^4 \, {\rm s}^{-1})$. 
Figure   2 
shows the total bulk viscosity 
$\zeta_0 = \zeta_{e0} + \zeta_{\mu 0} $ of non-superfluid matter 
versus nucleon density $n_b$. 
We have used models I and II of $npe\mu$ matter described 
in Sect.\ 2.3. The dotted lines show the bulk viscosity 
for the simplified models in which the 
muons are absent (cf.\ with Fig.\  1). 
The latter results coincide with those reported by Haensel 
\& Schaeffer (\cite{hs92}). 
 
The results for models I and II are similar. 
The bulk viscosity due to the direct Urca processes 
is switched on in a jump-like manner 
at the threshold density at which the electron direct 
Urca process becomes 
operative (this happens at 
$n_b$= 0.414 fm$^{-3}$ and 0.302 fm$^{-3}$, respectively). 
The presence of muons lowers the threshold 
density (mainly due to increasing the number density 
and Fermi momenta of protons). On the other hand, the 
muons lower the bulk viscosity produced by the electron direct Urca 
process (by decreasing $n_e$). At larger $n_b$, the total bulk 
viscosity suffers the second jump 
(at $n_b=$ 0.503 fm$^{-3}$ and 0.358 fm$^{-3}$ for models 
I and II, respectively). This time 
it is associated with switching on the muon direct 
Urca process, where muons participate by themselves. 
The contribution of the muon direct Urca into the bulk 
viscosity is even larger than the contribution of the 
electron direct Urca. The total bulk viscosity exceeds 
the bulk viscosity in $npe$ matter. This is 
natural: the muons introduce additional
non-equilibrium Urca process which makes
stellar matter more viscous.

Finally, let us discuss practical calculation of
the bulk viscosity. A user possesses usually
number densities of different particles for any given
equation of state of $npe\mu$ matter. This information
is sufficient to calculate factors $C_l$ numerically
from the second equality in Eq.\ (\ref{C_emu}). Other
density dependent 
quantities which enter the expressions for the
bulk viscosity are also expressed through the number densities.
Thus, the evaluation of the bulk viscosity for any equation
of state is not a problem.

Four our illustrative equations of state 
the functions $q_l(n_b)$=$
(n_l/n_0)^{1/3}$ $(C_l(n_b)/100\,{\rm MeV})^2$, which enter 
Eq.\ (\ref{zeta_NSF}), can be fitted by simple expressions 
\begin{equation} 
       q_l = a_0 + a_1\, u + a_2 \, u^2 + a_3 \, u^3 , 
\label{fit1} 
\end{equation} 
where $u=n_b/n_0 \leq 15$, $n_0=0.16$ fm$^{-3}$, 
and the maximum error $\la 0.5$\%. 
The fit parameters are: 
$a_0= -0.3849$, $a_1= 0.3338$, $a_2= 0.0375$, $a_3= -0.00098$ 
for $l=e$ and model I; 
$a_0= -0.3396$, $a_1= 0.4334$, $a_2= 0.0293$, $a_3= -0.0007$ 
for $l=\mu$ and model I; 
$a_0=-1.0491$, $a_1=1.1788$, $a_2=-0.013$, $a_3=0.00044$ 
for $l=e$ and model II; 
$a_0=-0.844$, $a_1=1.2715$, $a_2=-0.02023$, $a_3=0.000675$ 
for $l=\mu$ and model II. 
%
%
%
 
On the other hand, the baryon number number density 
as a function of mass density (for the model equations of state 
with the compression modulus $K_0=180$ MeV) can be fitted as 
\begin{equation} 
    n_b=n_0 \, b_1 \rho /( \rho_0+ b_2 \rho), 
\label{fit2} 
\end{equation} 
where $b_1=1.057$ and $b_2=0.0322$ for model I; 
$b_1=1.059$ and $b_2=0.0343$ for model II. 
The maximum fit error is $\la 0.9$\%, 
and $\rho_0=2.8 \times 10^{14}$ g cm$^{-3}$. 
%
%
%
These equations allow one to calculate the bulk viscosity 
as a function of mass density as required in practical
applications.

\section{Bulk viscosity of superfluid matter} 
 
\subsection{Superfluid gaps} 
 
Superfluidity of nucleons in a neutron star core
may strongly affect the bulk viscosity. 
Neutrons are believed to form Cooper 
pairs due to their interaction in the triplet 
state, while protons
suffer singlet-state pairing (Sect.\ 1). 
While studying the triplet-state neutron pairing 
one should distinguish the cases of 
different projections $m_J$ of $nn$-pair moment onto a 
quantization axis $z$ (see, e.g., Amundsen and {\O}stgaard \cite{ao85}):
$|m_J|=0,\, 1,\, 2$. The actual (energetically most 
favorable) state of $nn$-pairs 
is not known being extremely sensitive to the 
(still unknown) details of $nn$ interaction. 
One cannot exclude 
that this state varies with density and is a superposition 
of states with different $m_J$. 
We will consider 
the $^3$P$_2$-state neutron 
superfluidity either with $m_J=0$ or with $|m_J|=2$. 
In these two cases the effect of superfluidity on the bulk viscosity 
is qualitatively different. Consideration of the superfluidity
based on mixed $m_J$ states is much more complicated
and goes beyond the scope of the present paper.

Thus we will study three different superfluidity types: 
$^1$S$_0$, $^3$P$_2$ ($m_J=0$) and 
$^3$P$_2$ ($|m_J|=2$) denoted as A, B and C, 
respectively (Table \ref{tab:ABC}). 
The superfluidity of type A may be attributed to any 
protons, while superfluidity
of types B and C may be attributed to neutrons. 
 
Microscopically, superfluidity introduces 
an energy gap $\delta$ in momentum dependence of the nucleon energy,
$\varepsilon ( {\vec p} )$. Near the Fermi 
level ($ | p-\pF | \ll \pF$), this dependence can be written as 
\begin{equation} 
\begin{array}{l} 
      \varepsilon = \mu - \sqrt{\delta^2 + v_{\rm F}^2(p-p_{\rm F})^2} 
      \; \; {\rm at} \; \; p < \pF \, , 
      \\ 
      \varepsilon = \mu + \sqrt{\delta^2 + v_{\rm F}^2(p-p_{\rm F})^2} 
      \; \; {\rm at} \; \; p \ge \pF \; , 
\end{array} 
\label{Gap} 
\end{equation} 
where $\pF$ and $\vF$ are the Fermi 
momentum and Fermi velocity of the nucleon,
respectively, and $\mu $ is the 
nucleon chemical potential. In the cases of study
one has $\delta^2 = \Delta^2 (T) F(\vartheta)$, 
where $ \Delta(T)$ is the amplitude which describes 
temperature dependence of the gap; $F(\vartheta)$ 
specifies dependence of the gap on the angle $\vartheta$ between 
the particle momentum and the $z$ axis 
(Table \ref{tab:ABC}). 
In case A the gap is isotropic, and $\delta = \Delta(T)$. 
In cases B and C the gap depends on $\vartheta$. 
Note that in case C the gap vanishes at the poles 
of the Fermi sphere at any temperature: 
$ F_{\rm C} (0) = F_{\rm C} (\pi) = 0$. 
 
The gap amplitude $\Delta(T)$ is derived from the standard 
equation of the BCS theory 
(see, e.g., Yakovlev et al.\ \cite{yls99}).
The value of  $\Delta(0)$ determines 
the critical temperature $T_c$. 
The values of $k_{\rm B} T_c / \Delta(0)$ for cases A, B and C 
are given in Table \ref{tab:ABC}. 

\begin{table}[t] 
\renewcommand{\arraystretch}{1.2} 
\caption{Studied type of superfluidity} 
\begin{center} 
  \begin{tabular}{||c|ccc||} 
  \hline \hline 
Type\rule{0em}{2.5ex}
         & Pairing state &   $F(\vartheta)$ 
         & $\kB T_c/\Delta(0)$  \\ 
  \hline 
  A       & $^1{\rm S}_0$    &  1                     & 0.5669 
                                                        \rule{0em}{3ex} \\ 
  B       & $^3{\rm P}_2\ (m_J =0)$ 
                             & $(1+3\cos^2\vartheta)$ & 0.8416            \\ 
  C       & $^3{\rm P}_2\ (|m_J| =2)$ 
                             & $\sin^2 \vartheta$     & 0.4926 
                                        $\displaystyle\vph{a\over F_F}$  \\ 
  \hline \hline 
\end{tabular} 
\label{tab:ABC} 
\end{center} 
\end{table} 

For further analysis 
it is convenient to introduce the dimensionless quantities: 
\begin{equation} 
     v = \frac{\Delta(T)}{
 T}, \quad 
     \tau = \frac{T}{T_c}, \quad 
     y = \frac{\delta}{
 T}. 
\label{DimLess2} 
\end{equation} 
The dimensionless gap $y$ 
can be presented in the form: 
\begin{equation} 
     y_{\rm A} = v_{\rm A}, \quad y_{\rm B} = 
     v_{\rm B} \, \sqrt{1+3\cos ^2\vartheta}, 
     \quad 
     y_{\rm C} = v_{\rm C} \, \sin \vartheta.
\label{y} 
\end{equation} 
The dimensionless gap amplitude $v$ depends only on $\tau$. 
In case A the quantity $v$ 
coincides with the isotropic dimensionless gap, while 
in cases B and C it represents, respectively, the 
minimum and maximum gap (as a function of $\vartheta$) 
on the nucleon Fermi surface. 
The dependence of $v$ on $\tau$ can be fitted as 
(Levenfish and Yakovlev \cite{ly94}):
\begin{eqnarray} 
     v_{\rm A} & = &  \sqrt{1-\tau} \left( 1.456 - \frac{0.157}{\sqrt{\tau}} + 
             \frac{1.764}{\tau} \right),  \nonumber \\ 
     v_{\rm B} & = &  \sqrt{1-\tau} \left( 0.7893 + \frac{1.188}{\tau} 
              \right) , \nonumber \\ 
     v_{\rm C} & = &  \frac{ \sqrt{1-\tau^4} }{\tau} 
              \left( 2.030 - 0.4903 \tau ^4 + 0.1727 \tau ^8 \right). 
\label{v_fit} 
\end{eqnarray} 
The mean errors of these fits are $ \la 1 \% $ 
for all $\tau \le 1$. 
 
\subsection{Superfluid reduction factors} 
 
Now let us consider the effects of nucleon superfluidity on 
the bulk viscosity. 
The dynamics of superfluid is generally
much more complicated than the dynamics of ordinary
fluids. Even the motion of  
matter which consists of particles of one species
is described by the equations of two-fluid hydrodynamics
(normal and superfluid components), and viscous dissipation
of the normal component is determined by three coefficients
of the second (bulk) viscosity (Landau \& Lifshitz \cite{ll87}).
Our main assumption is that stellar pulsations 
represent fluid motion of the first-sound type 
(particularly, temperature variations are neglected) in which
all constituents of matter move with the same
hydrodynamical velocity. In this case the  hydrodynamical
equations reduce to the equation of one-fluid hydrodynamics
with one coefficient of the bulk viscosity ($\zeta=\zeta_2$ in 
the notation of Landau \& Lifshitz, 1987).   
 
We will see that superfluidity 
{\it reduces} the bulk viscosity due to the appearance 
of energy gaps in the nucleon dispersion relation, Eq.\ (\ref{Gap}). 
Quite generally, the bulk viscosity 
can be presented in the form  
\begin{equation} 
    \zeta= \sum_{l} \zeta_{l0}\, R_{l},
\label{zeta2} 
\end{equation} 
where $\zeta_{l0}$ is a partial bulk viscosity
of non-superfluid matter, Eq.\ (\ref{zeta_NSF}),
and $R_{l}$ is a factor which describes
reduction of the partial bulk viscosity by 
superfluidity of nucleons 1 and 2 involving into a
corresponding direct Urca process. If both nucleons, 1 and 2,
belong to non-superfluid component of matter, we 
have $R_{l}=1$ and reproduce the results of Sect.\ 2.
 
Thus the problem consists in calculating 
the reduction factors $R_{l}$. Each factor
depends generally on two parameters, $v_1$ and $v_2$, 
which are dimensionless gap amplitudes of nucleons 1 and 2
(and on the type of superfluidity of these nucleons).
Let us study the
effect of superfluidity on the partial bulk viscosity. 
For this purpose let us reconsider derivation 
of the bulk viscosity (Sect.\ 2.1). If 
all constituents of matter 
have the same macroscopic velocity, 
the superfluidity affects noticeably only 
the factor $\lambda_{l}$ in the expression for the
bulk viscosity, Eq.\ (\ref{zeta_NSF}). As seen from
Eq.\ (\ref{lambda0}), the main factor affected by 
the superfluidity in $\lambda_{l}$
is the integral $\Delta I$, Eq.\ (\ref{dI}), 
which describes 
the asymmetry of the lepton production and capture rates 
in the direct and inverse reactions of the direct Urca
process. 
At $\xi \ll 1$ the integrand of this equation is
$ J(x_\nu - \xi ) - J(x_\nu + \xi ) \approx 
-2\xi \: \partial J(x_\nu)/\partial x_\nu $, 
where $J(x_\nu)$ is given by 
Eq.\ (\ref{Jpm}). Thus, at small deviations from 
the equilibrium one can transform Eq.\ (\ref{dI}) to: 
\begin{eqnarray} 
   \Delta I_0 &= & 4 \xi 
       \int_0^{+\infty} \!\! {\rm d} x_\nu \, x_\nu 
       \!  \int_{-\infty}^{+\infty} \!\! {\rm d} x_1\, f(x_1) 
       \!  \int_{-\infty}^{+\infty} \!\! {\rm d} x_2\, f(x_2) 
\nonumber \\ 
       &  & \times 
       \!  \int_{-\infty}^{+\infty} \!\! {\rm d} x_e\, f(x_e)\, 
         \delta \left( x_1 + x_2 + x_e - x_\nu \right) . 
\label{I0} 
\end{eqnarray} 
Here, the index ``0'' refers to the non-superfluid case, in which 
we have obtained $\Delta I_0 = 17 \, \pi^4 \xi /60$. 
 
Generalization of $\lambda_{l}$
to the superfluid case can be achieved by introducing 
the neutron and proton energy gaps 
into Eq.\ (\ref{I0}). 
For convenience, let us define 
the dimensionless quantities 
\begin{equation} 
    x = \frac{v_{\rm F}(p-p_{\rm F})}{T}, \quad 
     z = \frac{\varepsilon - \mu}{T} 
     = {\rm sign}(x) \sqrt{x^2 + y^2}\, , 
\label{DimLess3} 
\end{equation} 
where $y$ is given by Eq.\ (\ref{DimLess2}). 
In the absence of superfluidity, we have $y = 0$ and 
$z=x$. 
 
Let the index $i=1$ correspond to 
a nucleon which can suffer superfluidity of type A while
$i=2$ correspond to a nucleon which can suffer any superfluidity,
A, B or C. 
In order to account for 
superfluidity in Eq.\ (\ref{I0}) it is sufficient to 
replace $x_i \to z_i$ for $i=1$ and 2 
[in $f(x_i)$ and in the delta
function] and introduce averaging
over orientations of ${\vec p}_2$  
(analogous procedure is considered in detail by
Levenfish \& Yakovlev \cite{ly94} for the problem 
of superfluid reduction of the neutrino emissivity).  
Then the  
factor $\lambda_{l}$
can be written as 
\begin{eqnarray} 
    \lambda_{l} & = & \lambda_{l0} \, R, \quad
    R(v_1,v_2)  = 
     \int_0^{\pi/2}\!\frac{{\rm d} \Omega}{4\pi} \; {\cal J}(y_1,y_2)
\nonumber \\ 
    & & = \int_0^{\pi/2}\!{\rm d}\vartheta \: 
    \sin \!\vartheta \; {\cal J}(y_1,y_2),
\label{Rdef} 
\end{eqnarray} 
where $\lambda_{l0}$ refers to the non-superfluid case,
$R=R_{l}$ is the reduction factor in question, and
\begin{eqnarray} 
      {\cal J}(y_1,y_2) & = &  \gamma
      \int_0^{+\infty} \!\!{\rm d} x_\nu \, x_\nu 
      \int_{-\infty}^{+\infty} \!\!{\rm d} x_l \, f(x_l) 
      \int_{-\infty}^{+\infty} \!\!{\rm d} x_1 \, f(z_l) 
\nonumber \\ 
     & & \times 
    \int_{-\infty}^{+\infty} \!\!{\rm d} x_2 \, f(z_2) \: 
          \delta( z_1 + z_2 + x_l - x_\nu) , 
\label{I} 
\end{eqnarray} 
with $\gamma=240/(17 \pi^4)$. 
Here, ${\rm d} \Omega$ is the solid angle element in
the direction of ${\vec p}_2$. 
 
Thus, we have derived explicit 
equations (\ref{Rdef}) and (\ref{I}) 
for calculating the reduction factor $R$. Calculation is quite 
similar (and in fact, simpler) to that 
done for the factor which describes superfluid reduction of 
the neutrino emissivity 
in the direct Urca process 
(Levenfish \& Yakovlev \cite{ly94}, Yakovlev et al.\ \cite{yls99}). 
The effect of superfluidity 
on the bulk viscosity has also much in common
with the effect on the emissivity. 
Thus we omit technical details and present only 
the results and their brief discussion. 
 
\subsection{Superfluidity of neutrons or protons}
 
Consider the superfluidity of nucleon of one species,
for instance, of species 2. 
In this case $R$ depends on the only parameter $v_2$, 
and we can set $z_1 = x_1$ in Eqs.\ (\ref{Rdef}) and (\ref{I}). 
Integration over $x_l$ and $x_1$ in Eq.\ (\ref{I}) 
reduces to well-known integrals of the theory of 
Fermi liquids and yields: 
\begin{eqnarray} 
     R & = & \gamma \int_0^{\pi/2} \!{\rm d} \vartheta \: 
        \sin\!\vartheta \int_0^{+\infty} {\rm d} x_\nu \, x_\nu 
        \int_0^{+\infty} {\rm d} x_2 
\nonumber \\	 
 && \times \, 
        \left[\, f(z_2) B(x_\nu -z_2) + f(-z_2)B(x_\nu + z_2) \, 
        \right], 
\label{Rone} 
\end{eqnarray} 
where $B(x) = x/({\rm e}^x -1)$. 
For $\tau =\tau_2 =T/T_{c2} \ge 1$, one has $R=1$. 
If superfluidity is strong 
($\tau \ll 1, \: v_2 \gg 1$), the direct Urca process is drastically 
suppressed by large superfluid gap in the nucleon spectrum 
and reduces the bulk viscosity. 
The asymptotic expressions of $R$ for $\tau \ll 1$ can be 
obtained from Eq.\ (\ref{Rone}): 
\begin{eqnarray} 
      R_{\rm A} & = & 
          \frac{20 \sqrt{2}}{17\pi^{3.5}} \: v^{3.5} \exp(-v) = 
          \frac{0.221}{\tau^{3.5}}   \; 
           \exp \left(- \frac{1.764}{\tau} \right), 
\label{Ra_asy} \\ 
     R_{\rm B} & = & 
           \frac{20\sqrt{3}}{51\pi^3} \; v^3 \exp(-v) = 
           \frac{0.0367}{\tau^3} \; 
           \exp \left(- \frac{1.188}{\tau} 
           \right), 
\label{Rb_asy} \\ 
      R_{\rm C}& = & 
           \frac{764 \pi^2}{1071} \: \frac{1}{v^2} = 1.7085 \,\tau^2 . 
\label{Rc_asy} 
\end{eqnarray} 
Note that the factors $R_{\rm A}$ and $R_{\rm B}$ 
are suppressed exponentially 
with decreasing temperature, whereas $R_{\rm C}$ varies as $T^2$. 
The latter fact is associated
with the presence of gap nodes at the Fermi surface
(Levenfish and Yakovlev \cite{ly94}).
 
\begin{figure}[t] 
\vspace*{-0.6cm} 
\begin{center} 
\epsfysize=9.5cm 
\epsfbox{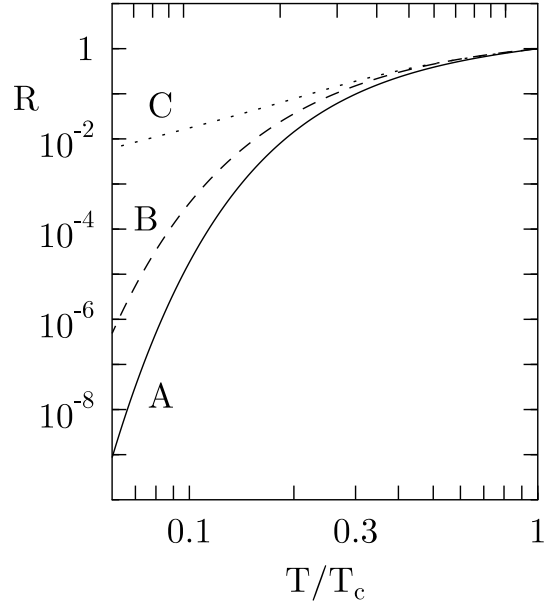} 
\vspace*{-0.65cm} 
\caption[]{\footnotesize 
    Reduction factors $R$ of the bulk viscosity 
    versus $\tau=T/T_c$ for three superfluidity types 
    A, B and C (Table \protect{\ref{tab:ABC}}). 
          } 
\end{center} 
\vspace*{-0.8cm}
\label{fig:Rone} 
\end{figure} 
 
In addition, we have calculated the reduction factors 
$R$ numerically in a wide range of $v$ 
and propose the expressions which fit 
the numerical results (with a mean error of $\la 1\%$) 
and reproduce the asymptotes (\ref{Ra_asy})--(\ref{Rc_asy}): 
\begin{eqnarray} 
 R_{\rm A} &=& 
         \left[0.2787 + \sqrt{ (0.7213)^2+(0.1564\,v)^2} \right]^{3.5}  
\nonumber      \\ 
       &  &\times \,
         \exp \left( 2.9965 - \sqrt{(2.9965)^2+v^2 } \right), 
\label{Rone_fitA} \\ 
 R_{\rm B}  &=& 
          \left[ 0.2854 + \sqrt{ (0.7146)^2+(0.1418\,v)^2} \right]^3 
\nonumber  \\ 
       &  & \times 
        \exp \left( 2.0350 - \sqrt{ (2.0350)^2+v^2 } \right), 
\label{eqLRone_fitB} \\ 
 R_{\rm C}  &=& 
          \frac{0.5+(0.1086\,v)^2}{1+(0.2347 \,v)^2+(0.2023 \,v)^4} 
\nonumber \\ 
         & & + 
        \, 0.5\, \exp\left( 1 - \sqrt{ 1+(0.5v)^2 } \right). 
\label{Rone_fitC} 
\end{eqnarray} 
Here, $v=v_{\rm A}$, $v=v_{\rm B}$ and $v=v_{\rm C}$ 
in the factors $R_{\rm A}$, $R_{\rm B}$ and $R_{\rm C}$, respectively. 
Using Eqs.\ (\ref{v_fit}) and 
(\ref{Rone_fitA})--(\ref{Rone_fitC}), 
one can easily calculate 
the reduction factors $R$ for any $\tau$. 
These factors are shown in Fig.\  3 
versus $\tau$. 
We see that the reduction can be quite substantial. 
The strongest reduction is provided by superfluidity A 
and the weakest by superfluidity C. For instance, at 
$T=0.1 \, T_c$ we obtain $R_{\rm A}\approx 2\times 10^{-5}$,  
$R_{\rm B}\approx 4\times 10^{-4}$ 
and $R_{\rm C}\approx 2\times 10^{-2}$.

\subsection{Superfluidity of neutrons and protons}
Let both nucleons, 1 and 2, be
superfluid at once, and let the superfluidity of 
nucleon 1 be of type A.
In this case $R$ can be calculated 
from Eqs.\ (\ref{Rdef}) and (\ref{I}). 
Using the delta function, we remove 
the integration over $x_\nu$ and obtain 
\begin{eqnarray} 
\lefteqn{
 {\cal J}(y_1,y_2)\! = 
                 \gamma \!\! 
                 \int_{-\infty}^{+\infty} \! \! \! {\rm d}x_1 \, f(z_1)\, 
   \!\!\! 
                 \int_{-\infty}^{+\infty} \! \! \! {\rm d}x_2 \, f(z_2) \, 
                 H(z_1+z_2), }
\label{Iboth} \\ 
\lefteqn{
       H(z) = \int_{-z}^{+\infty} 
       \! \! \! {\rm d}x \;\, \frac{z+x}{1+{\rm e}^x}. }
\label{H} 
\end{eqnarray} 
Notice that $H(z) \approx z^2/2$ as $z \to \infty$, and 
$H(z) \approx {\rm e}^z$ as $z \to - \infty$.

\begin{figure}[t] 
\begin{center} 
\epsfxsize=9cm 
\epsfbox{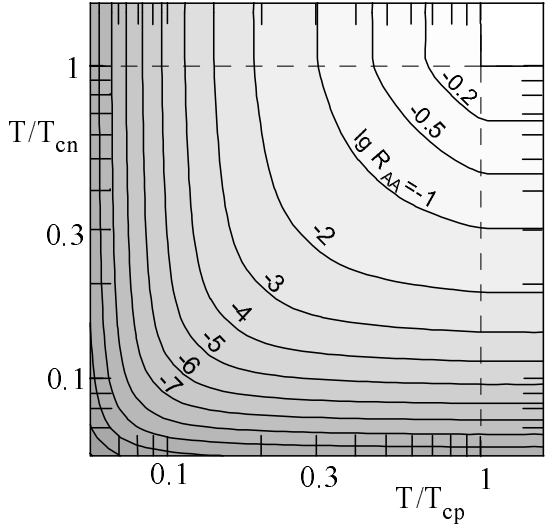} 
\vspace*{-4.2cm} 
\caption[]{\footnotesize 
    Isolevels of the reduction factor $R$ of the bulk viscosity
    by nucleon superfluidity of type AA
    versus $\tau_1=T/T_{c1}$ and $ \tau_2=T/T_{c2}$. 
    In the domain $\tau_1 \ge 1$, $\tau_2 \ge 1$ nucleons 1 and 2
    are normal and $R_{\rm AA}=1$. 
    In the domain $\tau_1 < 1$, $\tau_2 \ge 1$ 
    nucleons 1 are superfluid and nucleons 2 normal, while
    in the domain $\tau_1 \ge 1$, $\tau_2 < 1$ 
    nucleons 1 are normal and 2 superfluid;
    in these domains $R$ depends 
    on one parameter ($\tau_1$ or $\tau_2$). 
    In the domain $\tau_1 < 1$, $\tau_2 < 1$ both nucleons 1 and 2
    are superfluid at once. 
    } 
\end{center} 
\vspace*{-0.5cm} 
\label{fig:Raa} 
\end{figure} 

First consider the case in which the 
superfluidities of nucleons 1 and 2 are of type A.
Using Eqs.\ (\ref{y}) and (\ref{Rdef}) we get 
\begin{equation} 
   R_{\rm AA}(v_1,v_2) = {\cal J}(v_1,v_2) = {\cal J}(v_2,v_1), 
\label{Raa} 
\end{equation} 
where $v_1=y_1$ and $v_2=y_2$. 
It is evident that  $R_{\rm AA}(0,0)=1$. We have also derived 
the asymptote of $R_{\rm AA}$ in the limit of strong superfluidity. 
Furthermore, we have calculated the factor $R_{\rm AA}$ and derived 
the fit expression which reproduces the numerical 
results and the asymptotes. 
Both, the asymptotes and fits,
are given by the complicated expressions 
presented in the Appendix. 
In Fig.\  4 
we show the curves $R_{\rm AA}=$ const as a function 
of $\tau_1=T/T_{c1}$ and $ \tau_2=T/T_{c2}$. 
This visualizes the reduction the bulk viscosity 
for any $T$, $T_{c1}$ and $T_{c2}$. 
 
One can observe (Fig.\ 4) 
one important property of the reduction factor $R$. 
If both superfluidities are strong, 
$\tau_1^2 +\tau_2^2  \ll 1$, 
the factor $R$ is mainly determined by the 
larger of the two gaps (by the strongest 
superfluidity): 
\begin{equation} 
     R_{12} \sim \min \left\{ R_1,\, R_2 \vph{F\over F} \right\} . 
\label{Estim} 
\end{equation} 
Here, $R_1$ and $R_2$ are the reduction factors for 
the superfluidity of nucleons of one species.
The weaker superfluidity (with smaller energy gap) 
produces some additional reduction of the viscosity 
which is relatively small; this is confirmed 
by the asymptote $R_{\rm AA}$ given 
in the Appendix. The same effect takes place 
for the reduction of the neutrino emissivity 
(e.g., Yakovlev et al.\ \cite{yls99}).

\begin{figure}[t] 
\begin{center} 
\epsfxsize=9cm 
\epsfbox{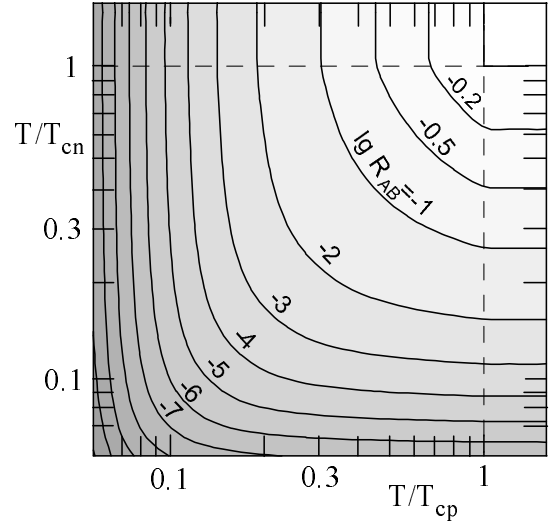} 
\vspace*{-4.5cm} 
\caption[]{\footnotesize 
    Same as in Fig.\  4 
    but 
    for the case in which the superfluidity of nucleons 1
    is of type A while the superfluidity of nucleons 2 is
    of type B. 
    } 
\end{center}     
\label{fig:Rab} 
\end{figure}

Equation (\ref{Rdef}) can be used also to evaluate 
$R$ for the case in which the nucleons of species 1
suffer pairing of type A, while the nucleons of species 2
suffer pairing of types B or C. 
In the Appendix we present the asymptotes of 
the factors $R_{\rm AB}$ and $R_{\rm AC}$  in the limit of strong 
superfluidity of both nucleon species
($\tau_1 \ll 1$, $\tau_2 \ll 1$). 
 
The factors $R_{\rm AB}$ and $R_{\rm AC}$ 
can be calculated easily 
in a wide range of $\tau_1$ and $\tau_2$. 
The calculation reduces to the one-dimensional integration in Eq.\ 
(\ref{Rdef}) of the function ${\cal J}(y_1,y_2)$ fitted 
by Eq.\ (\ref{Raa_fit}). 
The results are exhibited in Figs.\  5 
and  6. 
One can see that 
the dependence of the factors $R_{\rm AB}$ and $R_{\rm AC}$ 
on $\tau_1$ and $\tau_2$ 
has much in common with the dependence of $R_{\rm AA}$ but 
$R_{\rm AB}(\tau_1,\tau_2) \neq R_{\rm AB}(\tau_2,\tau_1)$ and 
$R_{\rm AC}(\tau_1,\tau_2) \neq R_{\rm AC}(\tau_2,\tau_1)$. The 
simple estimate (\ref{Estim}) turns out to be valid in 
cases AB and AC as well. 
However since the superfluidity of type C 
reduces the factor $R$ in a much weaker way 
than the superfluidities of types A or B, the 
transition from one dominating superfluidity 
to the other takes place in a rather wide region 
of $v_1$ and $v_2$ at $v_1 \sim \ln v_2$. 
Accordingly, for $v_2 \ga v_1 \gg 1$, 
the reduction factor 
$R_{\rm AC}$ exceeds greatly $R_{\rm AA}$ and $R_{\rm AB}$. 
 
For practical calculations of the bulk viscosity 
in superfluid matter, one needs to know how to 
evaluate $R_{\rm AA}$, $R_{\rm AB}$ and $R_{\rm AC}$. 
Corresponding expressions for superfluidity of one nucleon species
are given in Sect.\ 3.3. If nucleons 1 and 2 are superfluid at once,
the reduction factor $R_{\rm AA}$ can be determined easily from 
the fit equation (\ref{Raa_fit}). As for the reduction factors 
$R_{\rm AB}$ and $R_{\rm AC}$, we have generated their extensive tables.
These tables and 
numerical code which generates them are freely
distributed.
 
Finally, Fig.\ 7 
illustrates reduction 
of the bulk viscosity of $npe\mu$ matter 
with decreasing temperature 
by superfluidity of neutrons of type B or protons of type A 
for $n_b = 4 \, n_0$ and $\omega=10^4$ s$^{-1}$. 
Thick solid line shows the viscosity of non-superfluid matter 
(cf.\ with Fig.\ 2). 
Thin solid lines exhibit the bulk viscosity  
suppressed by the proton superfluidity at several 
selected critical temperatures $T_{cp}$ indicated near the curves. 
The dot-and-dashed line shows the effect of
neutron superfluidity ($T_{cn}= 10^{10}$ K) for normal protons. 
We see that the superfluid reduction of the bulk viscosity 
depends drastically on
temperature, superfluidity 
type, and critical temperatures $T_{cn}$ and $T_{cp}$. 
One can hardly expect $T_{cn}$ and $T_{cp}$ higher 
than $10^{10}$ K for $n_b$ as large as $4 \, n_0$ 
(e.g., Yakovlev et al.\ \cite{yls99}). If so, the superfluid
reduction cannot be very large, say, for $T \ga 3 \times 10^9$ K, 
but it can reach five orders of magnitude in 
the case of superfluid protons 
(or six orders of magnitude  
if $n$ and $p$ are superfluid at once,  
see Fig.\  5) 
for $T=10^9$ K at  
$T_{cn}=T_{cp}=10^{10}$ K.
It grows exponentially 
with further decrease of $T$. 
 
\begin{figure}[t] 
\begin{center} 
\epsfxsize=9cm 
\epsfbox{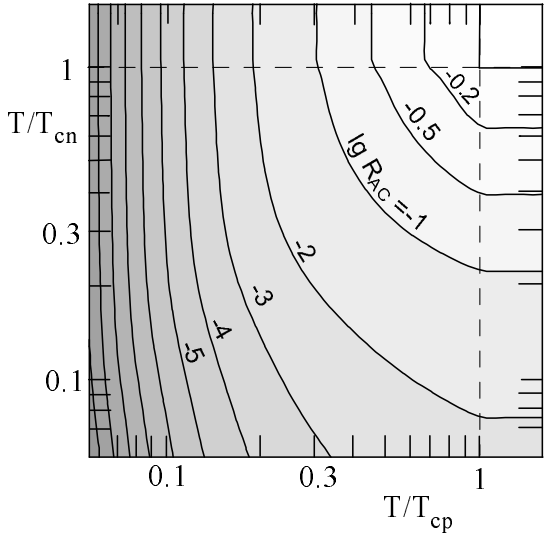} 
\vspace*{-4.5cm} 
\caption[]{\footnotesize 
    Same as in Fig.\  5 
    but for the case in which superfluidity 
    of nucleons 2 is of type C.
    } 
\end{center} 
\vspace*{-0.8cm}     
\label{fig:Rac} 
\end{figure} 
%
\begin{figure}[t] 
\vspace*{-2cm} 
\begin{center} 
\epsfxsize=9cm 
\epsfbox{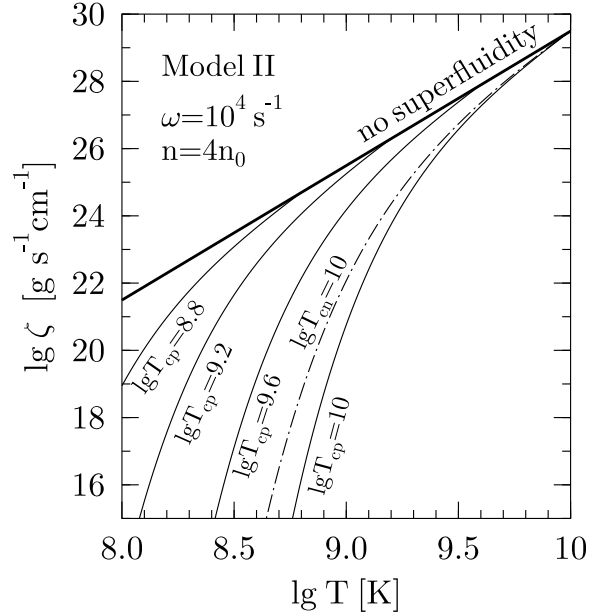} 
\vspace*{-0.5cm} 
\caption[]{\footnotesize 
 Bulk viscosity \protect{$\zeta$}  
 of superfluid $npe\mu$ matter (model II) 
 produced by the electron and muon direct Urca processes 
 at the baryon number density $n_b=4 \, n_0$ 
 and $\omega=10^4$ s$^{-1}$ as a function of temperature 
 for non-superfluid matter (thick solid line), for 
 matter with superfluid protons (solid curves, $T_{cp}=10^{10}$, 
 $10^{9.6}$, $10^{9.2}$ and $10^{8.8}$) 
 and normal neutrons, and for matter with superfluid 
 neutrons (dash-and-dotted curve, $T_{cn}=10^{10}$ K) and normal protons. 
} 
\end{center} 
\vspace*{-1cm} 
\label{fig:zeta-sup} 
\end{figure} 
 
\section{Conclusions} 
 
We have derived practical expressions for 
the bulk viscosity of matter in the cores of neutron 
stars under conditions in which the bulk viscosity is 
determined by
the direct Urca 
processes. We have paid special attention to the 
case in which dense matter consists of neutrons, 
protons, electrons and muons (Sects.\ 2.1--2.4).
In addition, we have studied the reduction of the bulk 
viscosity by superfluidity of neutrons and protons
(Sect.\ 3). We have analyzed the cases
of singlet-state superfluidity of protons,
and triplet-state superfluidity of neutrons (without 
and with the nodes of superfluid gaps on the nucleon 
Fermi surface). These cases are most interesting 
for applications (Sect.\ 1). We have 
obtained the practical expressions
for the bulk viscosity 
of superfluid $npe\mu$ matter.
The results can be 
used for studying the damping of neutron star pulsations
(Sect.\ 1) and in the studies of
the gravitational radiation driven instabilities 
in rotating neutron stars.
We will analyze the bulk viscosity induced by the
weaker modified Urca processes (\ref{baryon-Murca}) 
in a separate paper. 
 
\begin{acknowledgements}
We are grateful to the referee for very
useful critical remarks.
Two of the authors (KPL and DGY) acknowledge 
hospitality of N.\ Copernicus Astronomical 
Center in Warsaw. 
This work was supported in part by the 
RBRF (grant No. 99-02-18099), INTAS (grant No. 96-0542), 
and KBN (grant 2 P03D 014 13). 
\end{acknowledgements} 
 
\renewcommand{\theequation}{A\arabic{equation}} 
\setcounter{equation}{0} 
\section*{Appendix} 
Consider the case in which both nucleon species
are superfluid at once. 
The asymptotic behaviour of 
the reduction factor $R$ given by Eq.\ (\ref{Iboth}) 
in the limit of strong superfluidity 
depends on the gap amplitudes $v_1$ and $v_2$ 
and on the type of superfluidities. 
 
\subsection*{Case \protect{\rm AA}} 
 
Let both nucleon superfluidities be of type A.
If the superfluidities are strong ($v_1,v_2 \gg 1$) and 
$(v_1-v_2) \gg \sqrt{v_1} $ we obtain: 
\begin{eqnarray} 
    R_{\rm AA} & = & \gamma \: \sqrt{\frac{\pi v_1}{2}} \; 
            {\rm e}^{-v_1} \, K , 
\label{Raa_asy}\\ 
     K & = & \frac{s}{6} \, (v_1^2 + 2v_2^2) -\frac{1}{2}v_1 v_2^2\, \ln 
        \left(  \frac{v_1 + s }{v_2} \right) , 
\label{K} 
\end{eqnarray} 
where $s = \sqrt{|v_1^2 - v_2^2|}$. 
In the limiting case $v_2 \ll v_1$ 
Eq.\ (\ref{K}) yields  $K=v_1^3/60$ and 
reproduces  the asymptote given by Eq.\ 
(\ref{Ra_asy}). In the opposite limit 
$\sqrt{v_1} \ll (v_1 - v_2) \ll v_1$ one has $K=s^5/(15v_1^2)$. 
In the intermediate 
region $|v_2-v_1| \la \sqrt{v_2}$ the asymptote (\ref{K}) 
becomes invalid. One can show that 
$K \sim \sqrt{v_1}$  for $v_1 =v_2 $. 
 
We have calculated $R_{\rm AA}$ 
in a wide range of arguments $v_1$ and $v_2$ and 
proposed the fit expression valid for 
$\sqrt{v_1^2 + v_2^2} \la 50 $: 
\begin{equation} 
    R_{\rm AA} = \frac{u}{u+13.528}\: S \,+ \,D  . 
\label{Raa_fit} 
\end{equation} 
Here, 
\begin{eqnarray} 
  &&  
\begin{array}{l} 
\displaystyle   
  S= 
      \gamma \: 
      \left( K_0 + 2.3190\, K_1 + 0.016689 \, K_2 \vph{F\over F} \right)\: 
     \left( \frac{\pi}{2} \right)^{1/2} 
     p_s^{1/4} 
     \\ \quad  \quad \times \,
     \exp\left(-\sqrt{p_e} \right) , 
\end{array}      
\nonumber \\ 
 &&   K_0 = \displaystyle 
           \frac{\sqrt{p - q}}{6}\; ( p + 2 q) 
          - \sqrt{p}\, \frac{q}{2} \: 
          \ln \left( \frac{\sqrt{p}+\sqrt{p-q}}{\sqrt{q}} \right) , 
\nonumber \\ 
 &&      K_1 = \displaystyle \frac{\pi^2}{6} \sqrt{p - q} , 
\nonumber \\ 
 &&      K_2 = \displaystyle 
      \frac{p}{\sqrt{p - q}}\: \left( 1+\frac{\pi^2}{6} \right) , 
\nonumber \\ 
 &&       2p = u + 10.397 + \sqrt{w^2 + 5.4188\,u + 7.2987 } , 
\nonumber \\ 
 &&       2q = u + 10.397 - \sqrt{w^2 + 5.4188\,u + 7.2987  } , 
\nonumber \\ 
 &&      2p_s = u + \sqrt{w^2 + 3748.6\, u +  39.797 } , 
\nonumber \\ 
 &&      2p_e = u+  0.39417 + \sqrt{w^2 + 7.8849\, u + 1.8132 } , 
\nonumber \\ 
 &&     D=0.0080487\,  \left( q_1\,q_2\right)^{3/2} \, 
        \exp (- q_1 - q_2 + 9.9798 ) 
         , \mbox{\rule{0cm}{3ex}} 
\nonumber \\ 
 &&       q_1= 2.79205 + \sqrt{v_1^2 + (2.19785)^2} , 
\nonumber \\ 
 &&       q_2= 2.79205 + \sqrt{v_2^2 + (2.19785)^2} , 
\label{Raa_fit1} 
\end{eqnarray} 
with $ u=v_1^2+v_2^2$ and $w=v_1^2-v_2^2$.
 
\subsection*{Case \protect{\rm AB}} 
 
Let nucleons of species 1 suffer pairing of type A
and nucleons of species 2 suffer pairing of type B.
In this case $y_{\rm B}=y_2=v_2 \sqrt{1+3 \cos \vartheta^2}$, 
see Eq.\ (\ref{y}). 
There are three domains in the $(v_1,v_2)$-plane, 
in which the asymptotes of the reduction factor 
$R_{\rm AB}(v_1,v_2)$ in the limit of strong superfluidity 
are different. 
 
The first domain corresponds to $v_2>v_1$, i.e., to 
$y_2 > y_1$ for all $\vartheta$. 
If the both superfluidities are strong ($v_1 \gg 1$ and $v_2 \gg 1$) and 
$(v_2-v_1) \gg \sqrt{v_2}$, the asymptote of $R_{\rm AB}$ 
can be obtained by averaging  
Eq.\ (\ref{Raa_asy}) over $\vartheta$ 
after making a formal replacement 
$v_1 \to y_2$ and $v_2 \to v_1$. Since $v_2 \gg 1$ the main 
contribution into the integral comes from 
the region in which $|\cos \vartheta | \ll 1$. 
This allows us to put $y_2 \approx v_2 $ in all smooth 
functions under the integral. In this way we obtain 
\begin{eqnarray} 
  R_{\rm AB} & = & \gamma \; \frac{\pi}{2\sqrt{3}} \; 
           \exp (-v_2) 
\nonumber \\ 
    &&  \times \; 
           \left[ \,\frac{s}{6} \, (v_1^2 + 2v_2^2) -\frac{1}{2}v_1 v_2^2\, 
                 \ln\left(  \frac{v_1 + s }{v_2} \right) \, 
           \right] . 
\label{Rab_asy1} 
\end{eqnarray} 

The second domain corresponds to $v_1/2 > v_2$. We have 
$y_1>y_2$ for all $\vartheta$ 
in this domain. Then the asymptote of $R_{\rm AB}$ is derived 
by direct averaging over $\vartheta$ of the asymptote 
$R_{\rm AA}$ given by Eq.\ (\ref{Raa_asy}), 
after replacing formally $v_2 \to y_2$. We have: 
\begin{eqnarray} 
   R_{\rm AB} &=& 
	   \gamma \sqrt{ \frac{\pi}{2} \, v_1} \, 
                        \exp(-v_1) \, 
                        \left[ \frac{t}{24} \, 
                        \left( 3v_1^2+11v_2^2 \right)  \right. 
\nonumber  \\ 
           &  & + 
		   \frac{2\,v_1v_2^2}{6 \sqrt{3}} \arcsin 
                            \left( \frac{ \sqrt{3}\, v_1}{2s} \right)  
                     -   v_1v_2^2 \, 
                        \ln \left( \frac{v_1+t}{2v_2} \right) 
\nonumber \\ 
          &  &+   \left.      \frac{v_1^4-6v_1^2v_2^2-3v_2^4} 
                        {24 \sqrt{3}\,v_2} \, \arcsin 
                             \left(\frac{ \sqrt{3}\, v_2}{s} \right)  
                     \right] , 
\label{Rab_asy2} 
\end{eqnarray} 
where $t = \sqrt{v_1^2-4v_2^2}$. 
 
The third domain corresponds to 
$v_1/2 <v_2 <v_1$. While averaging over $\vartheta$ 
we can split the integral 
into two terms. The first term represents integration 
over the region 
$\cos \vartheta_0 <|\cos \vartheta| <1$, 
in which $y_1 < y_2$, while the second term contains integration 
over the region $0\le |\cos \vartheta| \le \cos \vartheta_0$, in which 
$y_1>y_2$, with 
$\cos \vartheta_0^2 =(v_1^2-v_2^2)/(3v_2^2)$. 
The latter term can be taken as an asymptote 
of the factor $R_{\rm AB}$, since the 
main contribution into the integral comes, again, 
from $|\cos \vartheta| \ll 1$. 
We have 
\begin{equation} 
      R_{\rm AB} = \gamma \sqrt{\frac{\pi\, v_1}{2}} \, 
           \exp(-v_1) \,  \frac{ \pi \, (v_1-v_2)^3 (v_1+3v_2)} 
           { 48 \sqrt{3} \,v_2}\, . 
\label{Rab_asy3} 
\end{equation} 
Note that the 
asymptotes given by Eqs.\ (\ref{Rab_asy1}) and (\ref{Rab_asy3}) 
are invalid  at $|v_2-v_1| \la \sqrt{v_1}$. 
If $v_2 \ll v_1$, Eq.\ (\ref{Rab_asy1}) reproduces Eq.\ 
(\ref{Ra_asy}) for $R_{\rm A}$.
For $v_1 \ll v_2$, Eq.\ (\ref{Rab_asy1}) 
transforms into Eq.\ (\ref{Rab_asy2}). 
Equations (\ref{Rab_asy2}) and 
(\ref{Rab_asy3}) coincide at $v_1 = 2v_2$. If $v_1 = v_2$, 
the reduction factor can be estimated as 
an $R_{\rm AB} \sim v^2 {\rm e}^{-v}$. 
 
\subsection*{Case \protect{\rm AC}} 
 
Let superfluidity of nucleons 1 be of type A, as before,
while superfluidity of nucleons 2 be of type C.
In this case  $y_{\rm C}=y_2 = v_2 \sin \vartheta$. 
For deriving $R_{\rm AC}$ in the limit of strong superfluidity we 
will use the asymptote of the function $H$ 
defined by Eq.\ (\ref{H}). 
If $v_1 > v_2\gg 1$, 
the main contribution into the integral 
(\ref{Iboth}) comes from 
the region, 
in which $z_1>z_2$. In this case the asymptote is 
\begin{eqnarray} 
   R_{\rm AC} & = & \gamma \sqrt{\frac{\pi v_1}{2}}\:\exp (-v_1) 
                   \left[ 
                  \frac{v_1}{8}(v_1^2 + v_2^2) 
                   - \, \frac{v_1v_2^2}{3} \: 
                   \ln \left( \frac{s}{v_2}   \right)  \right. 
\nonumber \\ 
          & &  \left.  +  \,\frac{v_1^4 -6v_1^2v_2^2 -3v_2^4}{24v_2}\: 
   \ln \left( \frac{v_1+v_2}{s}  \right)  
   \right] . 
\label{Rac_asy1} 
\end{eqnarray} 
If $v_2 \ll v_1 $, this asymptote reproduces 
the asymptote (\ref{Ra_asy}) for $R_{\rm A}$. For $v_1=v_2$, we have
$R_{\rm AC} \approx\gamma\sqrt{\pi v/2} \, {\rm e}^{-v} (3-4\ln 2)\, 
v^3/12$. 
Finally, at $v_2 >v_1 \gg 1$ 
the asymptote of $R_{\rm AC}$ is 
\begin{equation} 
   R_{\rm AC} =\gamma \sqrt{\frac{\pi v_1}{2}} \: \exp(-v_1) \: 
   \frac{v_1^5}{5v_2^2}\, . 
\label{Rac_asy2} 
\end{equation} 


\begin{thebibliography}{222} 
 
\bibitem[1985]{ao85} 
   Amundsen~L., {\O}stgaard~E., 1985, 
   Nucl.\ Phys.,  A442, 163 
 

\bibitem[1991]{bp91}
   Baym~G., Pethick~C.J., 1991, Landau Fermi-Liquid Theory,
   Wiley, New York

\bibitem[1990]{cls90} 
   Cutler~C., Lindblom~L., Splinter~R.J., 1990, ApJ, 363, 603 

\bibitem[1979]{fi79} 
   Flowers~E., Itoh~N., 1979, ApJ, 230, 847 

\bibitem[1979]{fm79} 
   Friman~B.L., Maxwell~O.V., 1979, ApJ, 232, 541 
 
\bibitem[1992]{hs92} 
   Haensel~P., Schaeffer~R., 1992, Phys.\ Rev., D45, 4708 
 

\bibitem[1987]{ll87}
   Landau~L.D., Lifshitz~E.M., 1987, Fluid Mechanics, Pergamon, Oxford 

\bibitem[1991]{lpph91} 
   Lattimer~J.M., Pethick~C.J., Prakash~M., Haensel~P., 
   1991, Phys.\ Rev.\ Lett., 66, 2701 
 
\bibitem[1995]{l95}
   Lindblom~L., 1995, ApJ, 438, 265

\bibitem[1995]{lm95}
   Lindblom~L., Mendel~G., 1995, ApJ, 444, 804

\bibitem[1998]{lom98}
   Lindblom~L. Owen~B.J., Morsink~S., 1998,
   Phys.\ Rev.\ Lett., 80, 4843

\bibitem[1994]{ly94} 
   Levenfish~K.P., Yakovlev~D.G., 1994, Astron.\ Lett., 20, 43 
 
\bibitem[1988]{mmst88} 
   Moeller~P., Myers~W.D., Swiatecki~W.J., Treiner~J. 1988, 
   Atomic Data Nucl.\ Data Tables, 39, 225

 
\bibitem[1988]{pal88}
   Prakash~M., Ainsworth~T.L., Lattimer~J.M., 1988,
   Phys.\ Rev.\ Lett., 61, 2518
 
 
\bibitem[1989]{s89} 
   Sawyer~R.F., 1989,  Phys.\ Rev., D39, 3804 
 
\bibitem[1983]{st83} 
   Shapiro~S.L., Teukolsky~S.A., 1983, 
   Black Holes  White Dwarfs  and Neutron Stars,
   Wiley-Interscience, New-York 
 
\bibitem[1995]{yl95} 
   Yakovlev~D.G., Levenfish~K.P., 1995, 
   A\&A, 297, 717 
 
\bibitem[1999]{yls99} 
   Yakovlev~D.G., Levenfish~K.P., Shibanov~Yu.A., 1999, 
   Physics-Uspekhi, 169, 825
 
\bibitem[1996]{jlz96}
    Zdunik~J.L., 1996, A\&A, 308, 828

\end{thebibliography}
\end{document}